\pdfoutput=1
\documentclass[aps,prb,10pt,twocolumn,superscriptaddress]{revtex4-2}
\usepackage{amsmath,amssymb,amsfonts,bm}
\usepackage{array,booktabs}
\usepackage{graphicx}
\usepackage{tikz}
\usepackage{xcolor-material}
\usepackage{hyperref}

\hypersetup{colorlinks}

\tikzset{
  x=1em, y=1em, node font=\footnotesize,
  vector box/.style={thick, MaterialTeal, fill=MaterialTeal100},
  tensor box/.style={thick, MaterialBlue, fill=MaterialBlue100},
  tensor leg/.style={semithick, MaterialGrey},
  covered tensor leg/.style={semithick, dashed, MaterialGrey300},
}
\usetikzlibrary{calc}

\newcommand{\ket}[1]{|#1\rangle}
\newcommand{\bra}[1]{\langle#1|}
\newcommand{\mel}[3]{\langle{#1}|{#2}|{#3}\rangle}
\newcommand{\1}{{\mathbf{1}}}
\newcommand{\ee}{{\mathrm{e}}}
\newcommand{\ii}{{\mathrm{i}}}

\DeclareMathOperator{\Tr}{Tr}

\newcommand{\Eprint}[2]{\href{#1}{arXiv:\allowbreak#2}}

\newcommand{\header}[2]{%
  \textsubscript{\footnotesize#1}\,\textbackslash\,\textsuperscript{\footnotesize#2}}

\newcommand{\Tetrahedron}[6]{
  \begin{tikzpicture}[baseline=1ex]
    \draw [covered tensor leg, MaterialGrey] (0,0) -- (3.2,0);
    \draw [tensor leg]
          (0,0) -- (2,2.5) -- (3.2,0) -- (2,-0.8) -- cycle
          (2,2.5) -- (2,-0.8);
    \draw (0.6, 1.7) node {$#1$}
          (3.2, 1.7) node {$#2$}
          (0.6,-0.9) node {$#3$}
          (3  ,-0.9) node {$#4$}
          (1.1, 0.5) node {$#5$}
          (2.4, 0.8) node {$#6$};
  \end{tikzpicture}
}
\newcommand{\Triangle}[6]{
  \begin{tikzpicture}[baseline=-0.5ex]
    \draw [tensor box]
          ( 90:1.5) -- (210:1.5) -- (330:1.5) -- cycle;
    \foreach \x in {90, 210, 330}
      \draw [tensor leg]
          (\x-40:1.7) .. controls (\x-20:0.8) and (\x+20:0.8) .. (\x+40:1.7)
          (\x-60:0.4) -- (\x-60:1.6);
    \draw ( 30:2) node {$#1$}
          (150:2) node {$#2$}
          (270:2) node {$#3$}
          ( 90:2) node {$#4$}
          (210:2) node {$#5$}
          (330:2) node {$#6$};
  \end{tikzpicture}
}

\hypersetup{
  pdftitle={Virasoro Generators in the Fibonacci Model Tensor Network --- Tackling Finite Size Effects},
  pdfauthor={Xiangdong Zeng, Ruoshui Wang, Ce Shen, Ling-Yan Hung}
}

\begin{document}

\title{Virasoro Generators in the Fibonacci Model Tensor Network --- \\ Tackling Finite Size Effects}
\author{Xiangdong Zeng}
\affiliation{State Key Laboratory of Surface Physics, Fudan University, Shanghai 200433, China}
\affiliation{Department of Physics and Center for Field Theory and Particle Physics, Fudan University, Shanghai 200433, China}
\author{Ruoshui Wang}
\affiliation{Cornell University, Ithaca, New York 14853, USA}
\affiliation{Institute for Advanced Study, Tsinghua University, Beijing 100084, China}
\author{Ce Shen}
\affiliation{State Key Laboratory of Surface Physics, Fudan University, Shanghai 200433, China}
\affiliation{Department of Physics and Center for Field Theory and Particle Physics, Fudan University, Shanghai 200433, China}
\author{Ling-Yan Hung}
\email{lyhung@fudan.edu.cn}
\affiliation{State Key Laboratory of Surface Physics, Fudan University, Shanghai 200433, China}
\affiliation{Department of Physics and Center for Field Theory and Particle Physics, Fudan University, Shanghai 200433, China}
\affiliation{Institute for Nanoelectronic devices and Quantum computing, Fudan University, Shanghai 200433, China}
\affiliation{Yau Mathematical Sciences Center, Tsinghua University, Beijing 100084, China}
\date{\today}

\begin{abstract}
  In this paper, we extend the method implementing Virasoro operators in a tensor network we proposed in \cite{Wang2022Virasoro} and test it on the Fibonacci model, which is known to suffer from far more finite size effects. To pick up the ``seed'' state that would flow to the stress tensor in the thermodynamic limit, we make use of the topological idempotent that projects the transfer matrix to the trivial sector. Combined with an optimization method, the seed state can be identified. We demonstrate that the descendant states in the Fibonacci model can be correctly generated with this approximate stress tensor, giving further evidence that the method applies more generally.
\end{abstract}

\maketitle

\section{Introduction}

There is a lot of progress in the last few years in recovering data of a continuous field theory from lattice models in a controlled numerical setting making use of tensor networks \cite{Evenbly2016Local,Hauru2016Topological,Milsted2017Extraction,Yang2017Loop,Bal2017Renormalization,Vanhove2018Mapping,Zou2018Conformal,Zou2020Conformal,Zou2020Emergence,Huang2022Numerical,Vanhove2022Critical}. Particularly, in the case of 2D conformal field theories (CFT), there is by now a lot of work on extracting the CFT spectra from the tensor networks. More recently, there are works developing methodologies to understand how conformal and other (continuous) symmetries are realized on the tensor networks \cite{Koo1994Representations,Milsted2017Extraction,Zou2018Conformal,Zou2020Conformal,Wang2022Emergence}.

These methods mainly focus on constructing discrete operators that act on a 1D lattice Hilbert space. However, they are not immediately applicable when we work on tensor networks that directly reconstruct the 2D partition functions. Frequently, we combine with tensor renormalization algorithms so that individual tensors building up the tensor network representation of the partition function are known only numerically, making it technically difficult to recover the Hamiltonian picture by taking the time evolution $\Delta t\to0$ limit. In \cite{Wang2022Virasoro}, we proposed a method to implement the Virasoro symmetries from the intuitive picture that the insertion of Virasoro operators is equivalent to the insertion of the stress tensor with appropriate weight around a closed loop. We implement stress tensor insertion by solving for the eigenstates of a the transfer matrix tiling a cylinder, which is related to the dilatation operator through a conformal map. By the state-operator correspondence, these eigenstates represent a small disk with the stress tensor inserted at its center. These small disks are then inserted around a large disk to extract Virasoro descendants of the state corresponding to the large disk. We demonstrated this method in the case of the Ising model and found that this naive method appears to work quite well.

The Ising model is known to be very special and suffers from less severe finite size effects. In this paper, we would like to inspect our proposal in other models where finite size effects are more severe. Particularly, we will consider the $\mathbb{Z}_3$ parafermion CFT, which can be obtained by giving appropriate boundary conditions to the Fibonacci Levin--Wen topological order. Here, the spectrum of a cylinder of intermediate sizes would still look quite far from the actual CFT spectrum. This however does not mean that we cannot extract anything useful out of it. As the cylinder gets bigger, the spectrum would approach that of the CFT, and there is a sense of continuity in the process of solving for eigenstates for cylinders of larger and larger sizes. It is well known that one can record the respective eigenvalues as cylinder size grows and this allows one to extract the asymptotic eigenvalues numerically with only cylinders of finite size. See for example \cite{Evenbly2016Local,Williamson2017Symmetry,Vanhove2018Mapping,Lootens2020Galois}.

In this paper, we will explain how we can pick up the eigenstate of a cylinder that would flow towards the stress tensor. A technical complication arises here, since the stress tensor is a fairly excited state in the spectrum, unlike in the case of the Ising model (we briefly reviewed the calculation of Ising CFT spectrum in Appendix~\ref{sec:ising-model}), making it more cumbersome to identify the eigenstate that would eventually flow to the actual stress tensor. To help identify the stress tensor, we make use of the projector constructed in \cite{Evenbly2016Local,Williamson2017Symmetry,Vanhove2018Mapping,Lootens2020Galois,Aasen2020Topological}. It can be constructed using the tube algebra method \cite{Evans1995Ocneanu,Evans1998Quantum}, and can project transfer matrix into sectors with definite topological charges. In the case of the stress tensor, it belongs to the trivial sector. Since the topological charges are explicit down to the smallest lattice scale in the lattice realization based on the string-net model, the eigenstate that could possibly flow to the stress tensor in the thermodynamic limit should still be found in the topologically trivial sector. The projector thus helps to remove a large number of states from other sectors. With the help of the projector, we identify the ``seed'' stress tensor which is used to construct the Virasoro operator that can be applied to a disk. We demonstrate that it generates descendants in the disk to an acceptable accuracy.

The paper is organized as follows. We will first review the tensor network construction of these lattice partition functions based on the Levin--Wen models. Then we will provide details of the transfer matrix of the Fibonacci model and construct the appropriate projector. Finally, we will demonstrate the construction of the Virasoro operator.

\section{Tensor network construction of lattice transfer matrices}

\subsection{Strange correlators}

It is observed that families of partition function $Z$ of some lattice models can be expressed as a \emph{strange correlator} \cite{Vanhove2018Mapping,Aasen2020Topological}, which is obtained by taking the overlap between the ground state wave-function of a topological order $\ket{\Psi}$ with some direct product state $\bra{\Omega}$. For concreteness, these topological orders are chosen to be those represented by the Levin--Wen string-net (SN) model, i.e.
\begin{equation}
  Z = \langle \Omega | \Psi_{\text{SN}} \rangle.
\end{equation}
When $\bra{\Omega}$ is tuned appropriately, the partition function would describe a 2D CFT in the thermodynamic limit.

Practically, the ground state wave-function of a given string-net model defined by a fusion category $\mathcal{C}$ can be constructed as a projected entangled pair states (PEPS) tensor network \cite{Buerschaper2009Explicit,Williamson2017Symmetry,Bultinck2017Anyons,Vanhove2018Mapping,Aasen2020Topological}. We will make heavy use of the notations introduced in \cite{Williamson2017Symmetry,Vanhove2018Mapping}. The general tensor unit of a trivalent PEPS graph is given by
\begin{multline}
  \Triangle jki\alpha\beta\gamma
  = D^{-2 (1/n_\alpha + 1/n_\beta + 1/n_\gamma)}
    \bigl( d_\alpha^{1/n_\alpha} d_\beta^{1/n_\beta} d_\gamma^{1/n_\gamma} \bigr) \\
    \cdot (d_i d_j d_k)^{-\frac14} (d_\alpha d_\beta d_\gamma)^{-\frac12}
    \Tetrahedron ik\gamma\alpha\beta j,
\end{multline}
where $d_i$ is the \emph{quantum dimension} of simple object $i$ belonging to $\mathcal{C}$, $D$ is the \emph{total quantum dimension} of $\mathcal{C}$ and $n_\alpha$ is the number of equal-length arcs in the closed loop containing $\alpha$. Each tetrahedron is assigned a value related to the $F$-symbols of the string-net model as follows:
\begin{equation}
  \Tetrahedron mjlnki = \sqrt{d_i d_j d_k d_l} \bigl[ F^{ijk}_l \bigr]_{mn}.
\end{equation}
The ground state wave-function $\ket{\Psi_{\text{SN}}}$, or the PEPS tensor network, is then constructed by assembling these building blocks, with the edge degrees of freedom playing the role of physical legs of the wave-function, and the corner degrees of freedom summed over:
\def\DiagramI{
  \begin{tikzpicture}[baseline=-0.5ex]
    \def\r{0.8}
    \def\d{0.2}
    \foreach \x in {30:1.5, 150:1.5, -90:1.5, -30:3}
      \draw [tensor box, shift=(\x)] (-30:1) -- (90:1) -- (-150:1) -- cycle;
    \foreach \x in {90:1.5, -30:1.5, -150:1.5, 30:3}
      \draw [tensor box, shift=(\x)] (30:1) -- (150:1) -- (-90:1) -- cycle;
    \begin{scope}[tensor leg]
      \draw (0,0) circle [radius=\r];
      \foreach \x/\xstart/\xdelta in {0/60/240, 60/180/180, 120/-120/120, 180/-60/120, -120/0/120, -60/0/180}
        \draw [shift=(\x:2.6)]
          ($(\xstart:\r) - cos(\xstart)*(0,\d) + sin(\xstart)*(\d,0)$) --
          (\xstart:\r) arc [radius=\r, start angle=\xstart, delta angle=\xdelta] --
          ++($cos(\xstart+\xdelta)*(0,\d) - sin(\xstart+\xdelta)*(\d,0)$);
      \foreach \x/\xstart in {30/180, -30/120}
        \draw [shift=(\x:4.5)]
          ($(\xstart:\r) - cos(\xstart)*(0,\d) + sin(\xstart)*(\d,0)$) --
          (\xstart:\r) arc [radius=\r, start angle=\xstart, delta angle=60] --
          ++($cos(\xstart+60)*(0,\d) - sin(\xstart+60)*(\d,0)$);
    \end{scope}
    \begin{scope}[semithick, MaterialTeal]
      \foreach \x in {0, 60, ..., 300}
        \draw [shift=(\x:1.3)]
          ($-sin(\x)*(0.45,0) + cos(\x)*(0,0.45)$) --
          ($ sin(\x)*(0.45,0) - cos(\x)*(0,0.45)$);
      \foreach \x in {30, -30}
        \draw [shift=(\x:2.25)]
          ($ cos(\x)*(0.45,0) + sin(\x)*(0,0.45)$) --
          ($-cos(\x)*(0.45,0) - sin(\x)*(0,0.45)$);
      \foreach \x in {90, 150, -150, -90}
        \draw [shift=(\x:1.5)]
          ($cos(\x)*(0.3 ,0) + sin(\x)*(0,0.3 )$) --
          ($cos(\x)*(0.95,0) + sin(\x)*(0,0.95)$);
      \foreach \x/\xstart in {90/30, -90/-30, -30/30, 30/-30}
        \draw [shift=(\xstart:3)]
          ($cos(\x)*(0.3 ,0) + sin(\x)*(0,0.3 )$) --
          ($cos(\x)*(0.95,0) + sin(\x)*(0,0.95)$);
    \end{scope}
  \end{tikzpicture}
}
\def\LinePhysical{
  \begin{tikzpicture}[baseline=1.6ex]
    \draw [semithick, MaterialTeal] (0,0) -- (0,2);
  \end{tikzpicture}
}
\def\LineVirtual{
  \begin{tikzpicture}[baseline=1.6ex]
    \draw [tensor leg] (0,0) -- (0,2);
  \end{tikzpicture}
}
\begin{equation}
  \DiagramI \enspace \text{where} \enspace
  \LinePhysical \, = \text{physical leg}, \enspace
  \LineVirtual  \, = \text{virtual leg}.
\end{equation}
By fixing the physical legs of the wave-function to some particular linear combinations (i.e.\ taking inner product with $\bra{\Omega}$), the strange correlator can be obtained.

Fibonacci string-net is defined on the hexagonal lattice. There are two objects, $\1$ and $\tau$ with quantum dimension $d_\1=1$, $d_\tau=\varphi$, where $\varphi=(1+\sqrt5)/2$ is the golden ratio. The only non-trivial fusion rule and $F$-symbol are
\begin{equation}
  \tau \times \tau = \1 + \tau, \quad
  F^{\tau\tau\tau}_\tau = \varphi^{-1} \begin{pmatrix} 1 & \sqrt{\varphi} \\ \sqrt{\varphi} & -1 \end{pmatrix}.
\end{equation}
The only non-vanishing tetrahedron and the corresponding triangle tensor are
\begin{equation}
  \begin{aligned}
       \Tetrahedron \tau mn\tau\tau\tau
    &= \sqrt{d_\tau d_\tau d_\tau d_\tau} \bigl[ F^{\tau\tau\tau}_\tau \bigr]_{mn}
     = \varphi^2 \bigl[ F^{\tau\tau\tau}_\tau \bigr]_{mn}, \\
       \Triangle \tau\tau mn\tau\tau
    &= (d_\tau d_\tau d_m)^{-\frac14} (d_\tau d_\tau d_n)^{-\frac13} \Tetrahedron \tau mn\tau\tau\tau \\
    &= \varphi^{\frac56} d_m^{-\frac14} d_n^{-\frac13} \bigl[ F^{\tau\tau\tau}_\tau \bigr]_{mn}, \enspace
       m,n \in \{\1,\tau\}.
  \end{aligned}
\end{equation}
In our strange correlator construction, all the physical legs are fixed to $\tau$ (i.e.\ $m=\tau$). So there will be only two kinds of triangle tensor units:
\begin{equation}
  \begin{aligned}
       \Triangle \tau\tau\tau\tau\tau\tau
    &= \varphi^{\frac14} \bigl[ F^{\tau\tau\tau}_\tau \bigr]_{\tau\tau} = -\varphi^{-\frac34}, \\
       \Triangle \tau\tau\tau\1\tau\tau
    &= \varphi^{\frac{7}{12}} \bigl[ F^{\tau\tau\tau}_\tau \bigr]_{\tau\1} = \varphi^{\frac{1}{12}}.
  \end{aligned}
  \label{eq:fib-triangle}
\end{equation}

\subsection{Cylinder transfer matrices}

The tensor units of Fibonacci string-net \eqref{eq:fib-triangle} are triangles. To calculate the corresponding spectrum conveniently, we first build a 4-leg tensor (a square) with a pair of triangles:
\def\DiagramI{
  \begin{tikzpicture}[baseline=-0.5ex, rotate=-30]
    \draw [tensor box, xshift=-1.2em]
          (60:1.5) -- (180:1.5) -- (-60:1.5) -- cycle;
    \draw [tensor box, xshift=1.2em]
          (0:1.5) -- (120:1.5) -- (-120:1.5) -- cycle;
    \draw [tensor leg, xshift=-1.2em]
          (0:0.4) -- (0:2)
          (120:0.4) -- (120:1.6) (-120:0.4) -- (-120:1.6)
          (140:1.7) .. controls (160:0.8) and (-160:0.8) .. (-140:1.7);
    \draw [tensor leg, xshift=1.2em]
          (60:0.4) -- (60:1.6) (-60:0.4) -- (-60:1.6)
          (40:1.7) .. controls (20:0.8) and (-20:0.8) .. (-40:1.7);
    \draw [tensor leg]
          ($(-1.2,0) + ( 100:1.7)$) .. controls (-0.8, 0.24) and (0.8, 0.24) .. ($(1.2,0) + ( 80:1.7)$)
          ($(-1.2,0) + (-100:1.7)$) .. controls (-0.8,-0.24) and (0.8,-0.24) .. ($(1.2,0) + (-80:1.7)$);
    \draw (-3.5,0) node {$i$}
          ( 3.5,0) node {$k$}
          ( 0, -2) node {$j$}
          ( 0,  2) node {$l$};
  \end{tikzpicture}
}
\def\DiagramII{
  \begin{tikzpicture}[baseline=1.5ex]
    \draw [tensor box]
          (0,0) -- (0,2) -- (2,2) -- (2,0) -- cycle;
    \draw [tensor leg]
          (-0.6,0.6) .. controls (0.2,0.6) and (0.6,0.2) .. (0.6,-0.6)
          ( 2.6,0.6) .. controls (1.8,0.6) and (1.4,0.2) .. (1.4,-0.6)
          ( 2.6,1.4) .. controls (1.8,1.4) and (1.4,1.8) .. (1.4, 2.6)
          (-0.6,1.4) .. controls (0.2,1.4) and (0.6,1.8) .. (0.6, 2.6);
    \foreach \x/\xtext in {45/l, 135/i, 225/j, 315/k}
      \draw (\x:2) + (1,1) node {$\xtext$};
  \end{tikzpicture}
}
\begin{equation}
  A_{ijkl} = \DiagramI = \DiagramII.
  \label{eq:fib-square}
\end{equation}
Using such a tensor unit, we would like to construct the transfer matrix $M$, which is the discrete version of the path-integral of the CFT on a cylinder. Consider putting the basic tensor units together to form a ring:
\def\DiagramI{
  \begin{tikzpicture}[baseline=0.5ex]
    \foreach \x in {0, 2, 4, 8}
      \draw [tensor box]
          (\x,0) -- (\x+1,0) -- (\x+1,1) -- (\x,1) -- cycle;
    \draw [covered tensor leg]
          (-0.7, 1.5) -- (9.7, 1.5)
          (-0.7,-0.5) -- (9.7,-0.5);
    \draw [tensor leg]
          (0.7, 2) -- (0.7, 1.5) .. controls (0.7,0.8) and (0.8,0.7) .. ( 1.5, 0.7)
                                 .. controls (2.2,0.7) and (2.3,0.8) .. ( 2.3, 1.5) -- (2.3, 2)
          (2.7, 2) -- (2.7, 1.5) .. controls (2.7,0.8) and (2.8,0.7) .. ( 3.5, 0.7)
                                 .. controls (4.2,0.7) and (4.3,0.8) .. ( 4.3, 1.5) -- (4.3, 2)
          (4.7, 2) -- (4.7, 1.5) .. controls (4.7,0.8) and (4.8,0.7) .. ( 5.5, 0.7)
          (8.3, 2) -- (8.3, 1.5) .. controls (8.3,0.8) and (8.2,0.7) .. ( 7.5, 0.7)
          (0.7,-1) -- (0.7,-0.5) .. controls (0.7,0.2) and (0.8,0.3) .. ( 1.5, 0.3)
                                 .. controls (2.2,0.3) and (2.3,0.2) .. ( 2.3,-0.5) -- (2.3,-1)
          (2.7,-1) -- (2.7,-0.5) .. controls (2.7,0.2) and (2.8,0.3) .. ( 3.5, 0.3)
                                 .. controls (4.2,0.3) and (4.3,0.2) .. ( 4.3,-0.5) -- (4.3,-1)
          (4.7,-1) -- (4.7,-0.5) .. controls (4.7,0.2) and (4.8,0.3) .. ( 5.5, 0.3)
          (8.3,-1) -- (8.3,-0.5) .. controls (8.3,0.2) and (8.2,0.3) .. ( 7.5, 0.3)
          (0.3, 2) -- (0.3, 1.5) .. controls ( 0.3,0.8) and ( 0.2, 0.7) .. (-0.5, 0.7) -- (-0.7,0.7)
                                 .. controls (-1.5,0.7) and (-1.5, 1.5) .. (-0.7, 1.5)
          (0.3,-1) -- (0.3,-0.5) .. controls ( 0.3,0.2) and ( 0.2, 0.3) .. (-0.5, 0.3) -- (-0.7,0.3)
                                 .. controls (-1.5,0.3) and (-1.5,-0.5) .. (-0.7,-0.5)
          (8.7, 2) -- (8.7, 1.5) .. controls ( 8.7,0.8) and ( 8.8, 0.7) .. ( 9.5, 0.7) -- ( 9.7,0.7)
                                 .. controls (10.5,0.7) and (10.5, 1.5) .. ( 9.7, 1.5)
          (8.7,-1) -- (8.7,-0.5) .. controls ( 8.7,0.2) and ( 8.8, 0.3) .. ( 9.5, 0.3) -- ( 9.7,0.3)
                                 .. controls (10.5,0.3) and (10.5,-0.5) .. ( 9.7,-0.5)
          (6.5,0.5) node {$\cdots$};
    \foreach \x/\xtext in {-0.3/1, 1.5/2, 3.5/3, 7.7/n}
      \draw (\x, 2.2) node {$i_\xtext$}
            (\x,-1.2) node {$j_\xtext$};
  \end{tikzpicture}
}
\begin{align}
     \tilde{M}_{i_1 i_2 \cdots i_n, \, j_1 j_2 \cdots j_n}
  &= \sum_{\substack{i_1, i_2, \ldots, i_n \\ j_1, j_2, \ldots, j_n}}
     \prod_{\alpha=1}^n A_{i_\alpha j_\alpha j_{\alpha+1} i_{\alpha+1}} \notag \\
  &= \DiagramI.
  \label{eq:fib-cylinder}
\end{align}
Different from the square lattice case (e.g. typically in the 2D Ising model), the indices of $A_{ijkl}$ are placed at corners, so they should be connected carefully in the particular order and orientation during contractions. Furthermore, the above cylinder $\tilde{M}$ is still not sufficient for a correct transfer matrix that is equal only to the exponentiation of the Hermitian. If we stack these cylinders row by row, the result will be a slanted cylinder and an additional phase will appear in the final spectrum. By putting two copies of cylinders with opposite orientations together, i.e.\ taking
\begin{equation}
  M = \tilde{M}\tilde{M}^\dagger,
\end{equation}
this phase can be removed.

As discussed in \cite{Hauru2016Topological,Wang2022Virasoro}, we can obtain the CFT spectrum by diagonalizing $T\cdot M$, where $T$ is the translation operator. In the tensor network representation, $T$ is simply given by shifting the lattice by one site:
\def\DiagramI{
  \begin{tikzpicture}[baseline=3ex]
    \draw [tensor leg]
          (0,0) .. controls (0,2) and (10,1) .. (10,3)
          (6,2.5) node {$\cdots$}
          (8,0.5) node {$\cdots$};
    \foreach \x in {0, 2, 4, 8}
      \draw [tensor leg]
          (\x,3) .. controls (\x,1.5) and (\x+2,1.5) .. (\x+2,0);
    \foreach \x/\xtext in {0/1, 2/2, 4/3, 10/n}
      \draw (\x,3) node [above] {$i_\xtext$}
            (\x,0) node [below] {$j_\xtext$};
    \draw (8,3) node [above] {$i_{n-1}$}
          (6,0) node [below] {$j_4$};
  \end{tikzpicture}
}
\begin{equation}
  T_{i_1 i_2 \cdots i_n, \, j_1 j_2 \cdots j_n} = \DiagramI.
  \label{eq:translation-operator}
\end{equation}

\section{CFT spectrum of Fibonacci string-net}

\subsection{Calculation}

In the thermodynamic limit, the Fibonacci string-net (critical hard-hexagon model with $c=4/5$ \cite{Vanhove2018Mapping}) approaches the $\mathbb{Z}_3$ parafermion CFT, but not all topological defects of the continuum CFT ($\mathbb{Z}_3\otimes\mathsf{Fib}$) have matrix product operator (MPO) representations in this specific lattice model \cite{Vanhove2018Mapping}. Therefore, to get the ``empty'' torus partition function without non-trivial topological defects, we have to look at a transfer matrix with $n=3k$ sites. Correspondingly, we should use a 3-site translation operator (or $T^3$) as well.

The memory consumption for the above construction grows exponentially with cylinder size $n$. To practically perform the calculation, we need to store the transfer matrix in the form of LinearOperator \cite{Virtanen2020SciPy}. The basic idea is that retaining the whole tremendous matrix in the memory is unnecessary; instead, what we need is the matrix-vector multiplication of this transfer matrix as an ``operator''. Due to the nature of tensor network representation, it's not difficult to construct such a multiplication function. The details of the implementation are given in Appendix~\ref{sec:implementation-details}.

\subsection{Spectrum of the transfer matrix}

The spectrum data are listed in \autoref{tab:fib-spectrum}. Fibonacci model suffers from significant finite size effects and hence only cylinders with size $n\gtrsim20$ can give a relatively accurate result. For small cylinders, some excitation states may not even show up among the low-lying states. In principle, we can separate different states based on their scaling dimensions and conformal spins. But due to the finite size effects, not only are the eigenvalues far from their limiting CFT values, there are ubiquitous issues with level-crossing --- some supposedly highly excited states in the thermodynamic limit pick up small eigenvalues when the cylinder is small, which makes it difficult to distinguish them. Here, we assume that the scaling dimensions $\Delta$ will approach the thermodynamic limiting values as $\Delta=A+B/n$ \cite{Schuler2016Universal} and we optimize the fitting results by selecting eigenstates of the small cylinder that actually flow to the desired CFT conformal states (i.e.\ choose the point nearest to the fitting curve; see Appendix~\ref{sec:fitting-details} for details). The ``$\infty$'' column in \autoref{tab:fib-spectrum} is obtained by extrapolating cylinder size $n\to\infty$ (equal to $A$ in the above formula). To estimate the scaling dimension more accurately, we can fix the identity state $\ket{\phi_\1}$ and state $\ket{\phi_T}$ corresponding to the stress tensor such that $\Delta_\1=0$ and $\Delta_T=2$:
\begin{equation}
  \Delta_\alpha = \frac{2}{\log\lambda_T - \log\lambda_I} \bigl( \log\lambda_\alpha - \log\lambda_I \bigr),
\end{equation}
where $\lambda_\alpha$ are the corresponding eigenvalues. The results are listed in the ``Rescaled'' column and plotted in \autoref{fig:fib-spectrum}.

\def\DiagramI{%
  \def\a{0.25}%
  \def\b{0.4}%
  \,%
  \begin{tikzpicture}[baseline=-0.75ex, semithick]
    \fill (\a,0) -- (0,\a) -- (-\a,0) -- (0,-\a) -- cycle;
    \draw (\b,0) -- (0,\b) -- (-\b,0) -- (0,-\b) -- cycle;
  \end{tikzpicture}%
  \,%
}
\begin{figure}[ht]
  \centering
  \includegraphics[width=\linewidth]{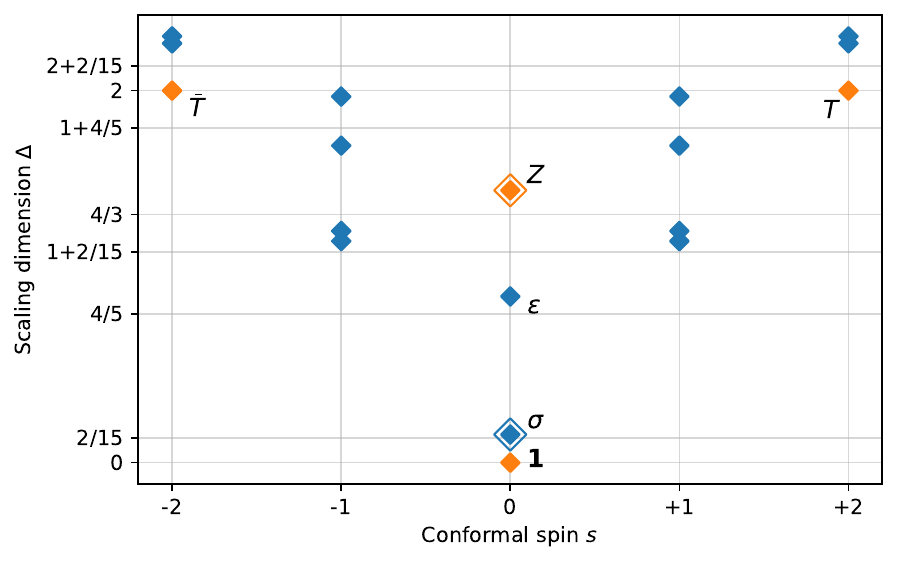}
  \caption{Fibonacci spectrum corresponding to the cylinder transfer matrix of size $n\to\infty$. ``\protect\DiagramI'' indicates that there is a 2-fold degeneracy. The values have been rescaled so that $\Delta_\1=0$ and $\Delta_T=2$. Primary states are labeled with their corresponding operators, with conformal dimensions $h_\1=0$, $h_\sigma=1/15$, $h_\epsilon=2/5$, $h_Z=2/3$, $h_X=7/5$ and $h_Y=3$, respectively (states corresponding to $X$ and $Y$ are higher and they are not considered in our calculation). The CFT partition function is then $Z=|\chi_\1+\chi_Y|^2+|\chi_\epsilon+\chi_X|^2+2|\chi_\sigma|^2+2|\chi_Z|^2$ \cite{Vanhove2018Mapping}. The states that lie within the sub-space corresponding to the trivial sector $|\chi_\1+\chi_Y|^2$ are drawn in orange.}
  \label{fig:fib-spectrum}
\end{figure}

\begin{table*}
  \centering\small
  \catcode`\"=\active
  \def"#1"{\textcolor{MaterialBlue700}{\bfseries#1}}
  \let\FM=\footnotemark
  \begin{tabular}{*{11}{c}}
    \toprule
      \header{Spin}{Size} &  9               &  12        &  15        &  18        &  21        &  24        &  27        &  $\infty$     &  Rescaled    &  Exact \\
    \midrule
      "0"                 & "0.0"            & "0.0"      & "0.0"      & "0.0"      & "0.0"      & "0.0"      & "0.0"      & "0.0"         & "0.0"        & "0"                       \\
       0\FM[1]            &  0.112899        &  0.113952  &  0.114474  &  0.114769  &  0.114950  &  0.115070  &  0.115152  &  0.11610\,(7) &  0.1513\,(32) &  $\frac{2}{15}$           \\
       0                  &  0.722756        &  0.709100  &  0.703083  &  0.699890  &  0.697989  &  0.696765  &  0.695931  &  0.6856\,(9)  &  0.894\,(19)  &  $\frac45$                \\
       $\pm1$             &  0.785553        &  0.817591  &  0.841528  &  0.859630  &  0.873686  &  0.884880  &  0.893989  &  0.9557\,(27) &  1.246\,(27)  &  $1{+}\frac{2}{15}$       \\
       $\pm1$             &  1.594955        &  1.346025  &  1.242196  &  1.184631  &  1.147942  &  1.122492  &  1.103796  &  0.914\,(11)  &  1.191\,(29)  &  $1{+}\frac{2}{15}$       \\
      "0"\FM[1]           & "1.290074"       & "1.222673" & "1.196184" & "1.182805" & "1.175050" & "1.170137" & "1.166820" & "1.123\,(4)"  & "1.464\,(32)" & "$\frac{\text4}{\text3}$" \\
       $\pm1$             &  1.147427        &  1.222649  &  1.274342  &  1.312081  &  1.340887  &  1.363620  &  1.382029  &  1.510\,(4)   &  1.97\,(4)    &  $1{+}\frac45$            \\
       $\pm1$             &  3.221770        &  2.490495  &  2.169235  &  2.016147  &  1.925379  &  1.865017  &  1.821876  &  1.31\,(5)    &  1.70\,(8)    &  $1{+}\frac45$            \\
      "\textpm2"          & "2.727358"\FM[2] & "2.139950" & "1.967143" & "1.887302" & "1.842910" & "1.815429" & "1.797149" & "1.535\,(33)" & "2.00\,(6)"   & "2"                       \\
       $\pm2$             &  1.032723\FM[2]  &  1.178227  &  1.276463  &  1.348394  &  1.403716  &  1.447739  &  1.483672  &  1.730\,(9)   &  2.25\,(5)    &  $2{+}\frac{2}{15}$       \\
       $\pm2$             &  -\FM[2]         &  1.430687  &  1.484488  &  1.527193  &  1.561317  &  1.589035  &  1.611935  &  1.759\,(9)   &  2.29\,(5)    &  $2{+}\frac{2}{15}$       \\
    \bottomrule
  \end{tabular}
  \footnotetext[1]{There is 2-fold degeneracy in these levels.}
  \footnotetext[2]{For $n=9$ cylinder, the transfer matrix is not large enough, so some high-level descendants may not show up; at the same time, spin $\pm2$ will degenerate to $\pm1$.}
  \caption{Fibonacci spectrum from transfer matrices of different sizes. Vacuum state and its descendants (marked with bold blue) are obtained by diagonalizing the transfer matrix with idempotents, see \autoref{subsec:topological-projectors}. The cylinder sizes are from 12 to 27 and are extrapolated to infinity, with data listed in the ``$\infty$'' column.}
  \label{tab:fib-spectrum}
\end{table*}

When there are many states mingled together, however, the optimization procedure is troublesome and may not be guaranteed to find the correct state. To simplify the task, we make use of the projector mentioned in the introduction. It was constructed in \cite{Bultinck2017Anyons,Williamson2017Symmetry,Vanhove2018Mapping,Lootens2019Cardy,Aasen2020Topological}, which would project the transfer matrix to a sub-space with definite topological charge. For the stress tensor, it belongs to the trivial sector, whose corresponding projector can be readily employed. In the next section, we will describe the construction of the needed idempotent in detail.

\subsection{Topological Projectors}
\label{subsec:topological-projectors}

In the PEPS representation of the string-net models, each type of anyons of the topological model has a corresponding string operator. Their construction is related to the so-called Ocneanu's tube algebra. Each type of anyon in the topological model corresponds to a topological sector of the CFT --- i.e.\ the families of primaries in the CFT can be organized into different topological sectors, each corresponding to an anyon of the topological model. Each string operator constructed in the topological model behaves as a projector that can project the CFT transfer matrix on a cylinder down to a sub-space containing states in this corresponding topological sector. Since we are particularly interested in obtaining the stress tensor, which belongs to the trivial sector in the topological model, we need the corresponding string operator. These operators take the form of matrix product operators (MPO) which have been systematically constructed in \cite{Bultinck2017Anyons,Williamson2017Symmetry,Vanhove2018Mapping,Lootens2019Cardy,Aasen2020Topological}, and they are often referred to as the \emph{idempotents}. We will review the needed ingredients below.

The tensor unit of such MPO is given by
\def\DiagramI{
  \begin{tikzpicture}[baseline=2.3ex]
    \draw [thick, MaterialOrange, fill=MaterialOrange100]
          (0,0) -- (0,2.4) -- (2.4,2.4) -- (2.4,0) -- cycle;
    \draw [tensor leg]
          (-0.6,0.6) .. controls (0.2,0.6) and (0.6,0.2) .. (0.6,-0.6)
          ( 3.0,0.6) .. controls (1.8,0.6) and (1.8,0.2) .. (1.8,-0.6)
          ( 3.0,1.8) .. controls (1.8,1.8) and (1.8,2.2) .. (1.8, 3.0)
          (-0.6,1.8) .. controls (0.2,1.8) and (0.6,2.2) .. (0.6, 3.0)
          (-0.6,1.2) -- (3,1.2) (1.2,-0.6) -- (1.2,3);
    \draw (-1.1,1.2) node {$j$}
          ( 3.5,1.2) node {$j$}
          (1.2,-1.2) node {$i$}
          (1.2, 3.6) node {$i$};
    \foreach \x/\xtext in {45/\delta, 135/\alpha, 225/\beta, 315/\gamma}
      \draw (\x:2.4) + (1.2,1.2) node {$\xtext$};
  \end{tikzpicture}
}
\def\DiagramII{
  \begin{tikzpicture}[baseline=2.3ex]
    \draw [thick, MaterialRed, fill=MaterialRed100]
          (0,0) -- (0,2.4) -- (2.4,2.4) -- (2.4,0) -- cycle;
    \draw [tensor leg]
          (-0.6,0.6) .. controls (0.2,0.6) and (0.6,0.2) .. (0.6,-0.6)
          ( 3.0,0.6) .. controls (1.8,0.6) and (1.8,0.2) .. (1.8,-0.6)
          ( 3.0,1.8) .. controls (1.8,1.8) and (1.8,2.2) .. (1.8, 3.0)
          (-0.6,1.8) .. controls (0.2,1.8) and (0.6,2.2) .. (0.6, 3.0)
          (-0.6,1.2) -- (3,1.2) (1.2,-0.6) -- (1.2,3) (0.7,1.7) -- (1.7,0.7);
    \draw (-1.1,1.2) node {$j$}
          ( 3.5,1.2) node {$j$}
          (1.2,-1.2) node {$i$}
          (1.2, 3.6) node {$i$}
          (2.5,-1.8) node {$k$};
    \foreach \x/\xtext in {45/\delta, 135/\alpha, 225/\beta, 315/\gamma}
      \draw (\x:2.4) + (1.2,1.2) node {$\xtext$};
    \draw [thin, MaterialGrey] (1.4,0.7) -- (2.5,-1.3);
  \end{tikzpicture}
}
\begin{equation}
  \begin{aligned}
       \DiagramI
    &= (d_\alpha d_\beta d_\gamma d_\delta)^{\frac14} G^{\beta i\gamma}_{j\alpha\delta}, \\
       \DiagramII
    &= (d_\alpha d_\beta d_\gamma d_\delta d_i d_j d_k)^{\frac14}
       G^{k\beta\delta}_{ij\alpha} G^{i\gamma\beta}_{kj\delta},
  \end{aligned}
\end{equation}
where $G$ is a normalized and symmetric version of the tetrahedron:
\begin{align}
     G^{abc}_{ijk}
  &= \frac{1}{\sqrt{d_j d_c}} \bigl[ F^{aik}_b \bigr]_{jc} \notag \\
  &= \frac{1}{\sqrt{d_a d_b d_c d_i d_j d_k}} \, \Tetrahedron jibcka.
\end{align}
For the cylinder transfer matrix of size $n$, the tube algebra basis is then given by a chain of $n-1$ orange tensors with a red one connected at the end (virtual legs to be contracted are omitted here):
\def\DiagramI{
  \begin{tikzpicture}[baseline=0.6ex]
    \foreach \x in {0, 2, 4}
      \draw [thick, MaterialOrange, fill=MaterialOrange100]
          (\x,0) -- (\x+1,0) -- (\x+1,1) -- (\x,1) -- cycle;
    \draw [thick, MaterialRed, fill=MaterialRed100]
          (8,0) -- (9,0) -- (9,1) -- (8,1) -- cycle;
    \foreach \x in {0, 2, 4}
      \draw [tensor leg] (\x+0.5,-1) -- (\x+0.5,2);
    \draw [tensor leg]
          (-1,0.5) -- (5.5,0.5) (7.5,0.5) -- (10,0.5)
          (8.5,-1) -- (8.5,2)
          (8.2,0.8) -- (8.8,0.2);
    \foreach \x in {0, 2, 4}
      \draw (\x+0.5,2) node [above] {$\tau$};
    \draw (-1.5, 0.5) node {$a$}
          (10.5, 0.5) node {$a$}
          ( 8.5, 2.5) node {$b$}
          ( 9.2,-0.5) node {$c$}
          ( 6.5, 0.5) node {$\cdots$};
  \end{tikzpicture}
}
\begin{equation}
  \mathcal{T}_{ab}^c = \DiagramI.
\end{equation}

In general, the idempotents are linear combinations of these basis tensors, and the calculation of their coefficients can be found in Appendix C of \cite{Bultinck2017Anyons}. Here, our aim is to identify the stress tensor which is the descendant of the vacuum state, so we need to use the idempotent projecting to the trivial sector. This is given by
\begin{equation}
  \mathcal{P}_1 = \frac{1}{\sqrt5}
  \left( \frac{1}{\phi} \mathcal{T}^\1_{\1\1} + \sqrt{\phi} \mathcal{T}^\tau_{\tau\1} \right).
\end{equation}
Then it can be stacked to the transfer matrix~\eqref{eq:fib-cylinder} with translation operator~\eqref{eq:translation-operator}. In \autoref{fig:fib-spectrum}, the states that lie within the sub-space preserved by $\mathcal{P}_1$ are drawn in orange. The stress tensor $T$ lies within this trivial sector, with many other intervening states with similar conformal dimensions removed. This allows us to recover the correct seed state that flows to the stress tensor.

\section{Virasoro operator}

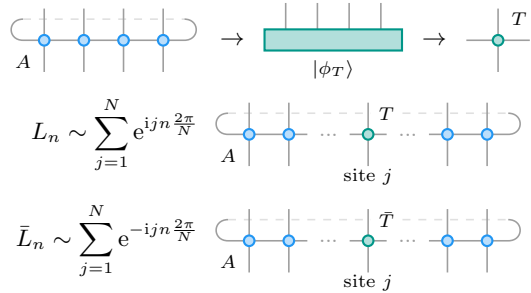
\begin{figure}[ht]
  \def\DiagramI{
  \begin{tikzpicture}[baseline=-0.5ex]
    \draw [tensor leg]
          (-0.8,0.8) .. controls (-1.4,0.8) and (-1.4,0) .. (-0.8,0) --
          ( 5.3,0) .. controls ( 5.9,0) and ( 5.9,0.8) .. ( 5.3,0.8);
    \draw [covered tensor leg] (-0.8,0.8) -- (5.3,0.8);
    \foreach \x in {0, 1, 2, 3} {
      \draw [tensor leg] (\x*1.5,-1.2) -- (\x*1.5,1.2);
      \draw [tensor box] (\x*1.5,0) circle [radius=0.2];
    }
    \draw (-0.8,-0.8) node {\scriptsize $A$};
  \end{tikzpicture}
}
\def\DiagramII{
  \begin{tikzpicture}[baseline=0.5ex]
    \foreach \x in {0, 1, 2, 3}
      \draw [tensor leg] (\x*1.2,0.8) -- ++(0,1);
    \draw [vector box] (-0.8,0) -- (4.4,0) -- (4.4,0.8) -- (-0.8,0.8) -- cycle;
    \draw (1.8,-0.7) node {\scriptsize $\ket{\phi_T}$};
  \end{tikzpicture}
}
\def\DiagramIII{
  \begin{tikzpicture}[baseline=-0.5ex]
    \draw [tensor leg] (-1.2,0) -- (1.2,0) (0,-1.2) -- (0,1.2);
    \draw [vector box] (0,0) circle [radius=0.2];
    \draw (0.8,0.8) node {\scriptsize $T$};
  \end{tikzpicture}
}
\def\DiagramIV#1{
  \begin{tikzpicture}[baseline=-0.5ex]
    \draw [tensor leg]
          (-0.8,0.8) .. controls (-1.4,0.8) and (-1.4,0) .. (-0.8,0) --
          (2.2,0) (3.8,0) -- (5.2,0) (6.8,0) --
          ( 9.8,0) .. controls (10.4,0) and (10.4,0.8) .. ( 9.8,0.8);
    \draw [covered tensor leg] (-0.8,0.8) -- (9.8,0.8);
    \foreach \x in {0, 1, 3, 5, 6}
      \draw [tensor leg] (\x*1.5,-1.2) -- (\x*1.5,1.2);
    \foreach \x in {0, 1, 5, 6}
      \draw [tensor box] (\x*1.5,0) circle [radius=0.2];
    \draw [vector box] (4.5,0) circle [radius=0.2];
    \draw (-0.8,-0.8) node {\scriptsize $A$}
          ( 5.2, 0.8) node [fill=white] {\scriptsize $#1$}
          (3,0) node [tensor leg] {...}
          (6,0) node [tensor leg] {...}
          (4.5,-1.6) node {\scriptsize site $j$};
  \end{tikzpicture}
}

\begin{gather*}
  \DiagramI \enspace \to \enspace \DiagramII \enspace \to \enspace \DiagramIII \\
  \begin{aligned}
    L_n       \sim \sum_{j=1}^N \ee^{ \ii j n \frac{2\pi}{N}} &\enspace \DiagramIV{T} \\
    \bar{L}_n \sim \sum_{j=1}^N \ee^{-\ii j n \frac{2\pi}{N}} &\enspace \DiagramIV{\bar{T}}
  \end{aligned}
\end{gather*}
  \caption{Tensor network implementation of equation~\eqref{eq:virasoro-operators}. Eigenstate $\ket{\phi_T}$ corresponding to the stress tensor is solved from the cylinder transfer matrix. It's reshaped to a 4-leg tensor and then inserted into a new cylinder to construct the Virasoro operators.}
  \label{fig:virasoro-construction}
\end{figure}

Our method for constructing the Virasoro operator in lattice models is inspired by the success of ``discrete holomorphicity'', a substantial subject reviewed for example in \cite{Cardy2009Discrete}. The basic idea is that a lattice version of contour integral can be approximated by discretely inserting operators along the intended path. In \cite{Wang2022Virasoro} we made the following proposal to implement Virasoro operators in a tensor network. We take the following definitions of the Virasoro operators in the continuous theory
\begin{equation}
  L_n       \sim \sum_{j=1}^N \ee^{ \ii j n \frac{2\pi}{N}} T(j), \quad
  \bar{L}_n \sim \sum_{j=1}^N \ee^{-\ii j n \frac{2\pi}{N}} \bar{T}(j).
  \label{eq:virasoro-operators}
\end{equation}
and take them seriously in a tensor network, by replacing the operator insertion at a point by the state we solved from the transfer matrix above at the vertex labeled $j$. Here, $T(j)$ and $\bar{T}(j)$ are eigenstates of the cylinder transfer matrix of finite size we found in previous sections. With the help of idempotents and optimization, they are deemed to flow towards the actual CFT states corresponding to the stress tensor in the thermodynamic limit. The insertion procedure is depicted in \autoref{fig:virasoro-construction}.

Since the cylinder's size have to be $n=3k$ for the Fibonacci case, we can't directly insert the stress tensor $T$ and $\bar{T}$ into the new cylinder as it requires a 4-leg unit. Instead, we need to first reshape it into a triangle and then pad with ``empty'' triangle tensors equation~\eqref{eq:fib-triangle} to obtain a square one, analogous to the tensor constructed in equation~\eqref{eq:fib-square}:
\def\DiagramI{
  \begin{tikzpicture}[baseline=1.2ex]
    \draw [vector box]
          (5,0.4) -- (0,0.4) -- (0,1) -- (5,1)
          (8,0.4) -- (9,0.4) -- (9,1) -- (8,1);
    \draw [tensor leg]
          ( 0.3, 1.5) .. controls (0.3,0.8) and (0.2,0.7) .. (-0.5, 0.7)
          ( 0.7, 1.5) .. controls (0.7,0.8) and (0.8,0.7) .. ( 1.5, 0.7)
                      .. controls (2.2,0.7) and (2.3,0.8) .. ( 2.3, 1.5)
          ( 2.7, 1.5) .. controls (2.7,0.8) and (2.8,0.7) .. ( 3.5, 0.7)
                      .. controls (4.2,0.7) and (4.3,0.8) .. ( 4.3, 1.5)
          ( 4.7, 1.5) .. controls (4.7,0.8) and (4.8,0.7) .. ( 5.5, 0.7)
          ( 8.3, 1.5) .. controls (8.3,0.8) and (8.2,0.7) .. ( 7.5, 0.7)
          ( 8.7, 1.5) .. controls (8.7,0.8) and (8.8,0.7) .. ( 9.5, 0.7)
          ( 6.5, 0.7) node {$\cdots$};
    \draw [covered tensor leg]
          (-1.5,0.7) -- (-0.5,0.7)
          (10.5,0.7) -- ( 9.5,0.7);
    \foreach \x/\xtext in {-1/1, 1.5/2, 3.5/3, 7/n=3k}
      \draw (\x, 1.6) node {$i_{\xtext}$};
  \end{tikzpicture}
}
\def\DiagramII{
  \begin{tikzpicture}[baseline=-1.8ex]
    \def\r{0.6}
    \def\d{0.2}
    \begin{scope}
      \draw [vector box] (30:2+\r) -- (150:2+\r) -- (-90:2+\r) -- cycle;
      \foreach \x/\xangle in {60/-180, 120/-180, 180/-60, -120/-60, -60/60, 0/60}
        \draw [tensor leg, shift=(\xangle:\r)]
          ($(\x:1.73) + cos(\xangle-90)*(\d,0) + sin(\xangle-90)*(0,\d)$) --
          (\x:1.73) arc [radius=\r, start angle=\xangle, delta angle=180] --
          ++($cos(\xangle-90)*(\d,0) + sin(\xangle-90)*(0,\d)$);
      \foreach \x in {30, 150, -90}
        \draw [tensor leg, shift=(\x+150:\r)]
          ($(\x:3) + cos(\x+60)*(\d,0) + sin(\x+60)*(0,\d)$) --
          (\x:3) arc [radius=\r, start angle=\x+150, delta angle=60] --
          ++($cos(\x-60)*(\d,0) + sin(\x-60)*(0,\d)$);
      \draw
            (-2.8, 1.8) node {$i_1$}
            (-2.2,-0.2) node {$i_2$}
            (-1.3,-1.6) node [rotate=120] {$\cdots$}
            ( 0  ,-3.2) node {$i_{k+1}$}
            ( 1.9,-1.7) node {$i_{k+2}$}
            ( 2.2,-0.3) node [rotate=-120] {$\cdots$}
            ( 3.6, 1.8) node {$i_{2k+1}$}
            ( 0.9, 2.2) node {$i_{2k+2}$}
            (-0.8, 2.2) node {$\cdots$};
    \end{scope}
  \end{tikzpicture}
}
\def\DiagramIII{
  \begin{tikzpicture}[baseline=-1.8ex, rotate=-30]
    \def\r{0.6}
    \def\d{0.2}
    \begin{scope}
      \draw [vector box] (60:2+\r) -- (180:2+\r) -- (-60:2+\r) -- cycle;
      \foreach \x/\xangle in {90/-150, 150/-150, -150/-30, -90/-30}
        \draw [tensor leg, shift=(\xangle:\r)]
          ($(\x:1.73) + cos(\xangle-90)*(\d,0) + sin(\xangle-90)*(0,\d)$) --
          (\x:1.73) arc [radius=\r, start angle=\xangle, delta angle=180] --
          ++($cos(\xangle-90)*(\d,0) + sin(\xangle-90)*(0,\d)$);
      \draw [tensor leg, shift=(-30:\r)]
        ($(180:3) + cos(120)*(\d,0) - sin(120)*(0,\d)$) --
        (180:3) arc [radius=\r, start angle=-30, delta angle=60] --
        ++($cos(120)*(\d,0) + sin(120)*(0,\d)$);
    \end{scope}
    \begin{scope}[xshift=3.2em]
      \foreach \x in {0:2, 120:2, -120:2, 60:1, 180:1, -60:1}
        \draw [tensor box, shift=(\x)] (0:\r) -- (120:\r) -- (-120:\r) -- cycle;
      \foreach \x in {0:1, 120:1, -120:1}
        \draw [tensor box, shift=(\x)] (60:\r) -- (180:\r) -- (-60:\r) -- cycle;
      \draw [tensor leg] (0,0) circle [radius=\r];
      \foreach \x/\xangle in {30/150, 90/150, -90/30, -30/30}
        \draw [tensor leg, shift=(\xangle:\r)]
          ($(\x:1.73) + cos(\xangle-90)*(\d,0) + sin(\xangle-90)*(0,\d)$) --
          (\x:1.73) arc [radius=\r, start angle=\xangle, delta angle=180] --
          ++($cos(\xangle-90)*(\d,0) + sin(\xangle-90)*(0,\d)$);
      \foreach \x/\xangle in {150/-90, -150/-90}
        \draw [tensor leg, shift=(\xangle:\r)]
          (\x:1.73) arc [radius=\r, start angle=\xangle, delta angle=180] --
          ++(-\d,0) arc [radius=\r, start angle=90, delta angle=180] -- cycle;
      \draw [tensor leg, shift=(150:\r)]
        ($(0:3) + cos(60)*(\d,0) + sin(60)*(0,\d)$) --
        (0:3) arc [radius=\r, start angle=150, delta angle=60] --
        ++($cos(60)*(\d,0) - sin(60)*(0,\d)$);
      \draw [tensor leg, shift=(-30:\r)]
        ($(120:3) + cos(60)*(\d,0) + sin(60)*(0,\d)$) --
        (120:3) arc [radius=\r, start angle=-30, delta angle=-60] --
        ++(-\d,0) arc [radius=\r, start angle=-90, delta angle=-60] --
        ++($-cos(60)*(\d,0) + sin(60)*(0,\d)$);
      \draw [tensor leg, shift=(30:\r)]
        ($(-120:3) + cos(60)*(\d,0) - sin(60)*(0,\d)$) --
        (-120:3) arc [radius=\r, start angle=30, delta angle=60] --
        ++(-\d,0) arc [radius=\r, start angle=90, delta angle=60] --
        ++($-cos(60)*(\d,0) - sin(60)*(0,\d)$);
    \end{scope}
  \end{tikzpicture}
}
\begin{align}
  \DiagramI &=   \DiagramII \notag \\
            &\to \DiagramIII.
\end{align}
Note that the contraction only involves the corner degrees of freedom. For the other sites of the new cylinder, we can build them in the same way. To make the contraction more efficient, however, we can block $k\times k$ square tensors in equation~\eqref{eq:fib-square} as the unit equivalently.

\begin{figure}[ht]
  \centering
  \includegraphics[width=\linewidth]{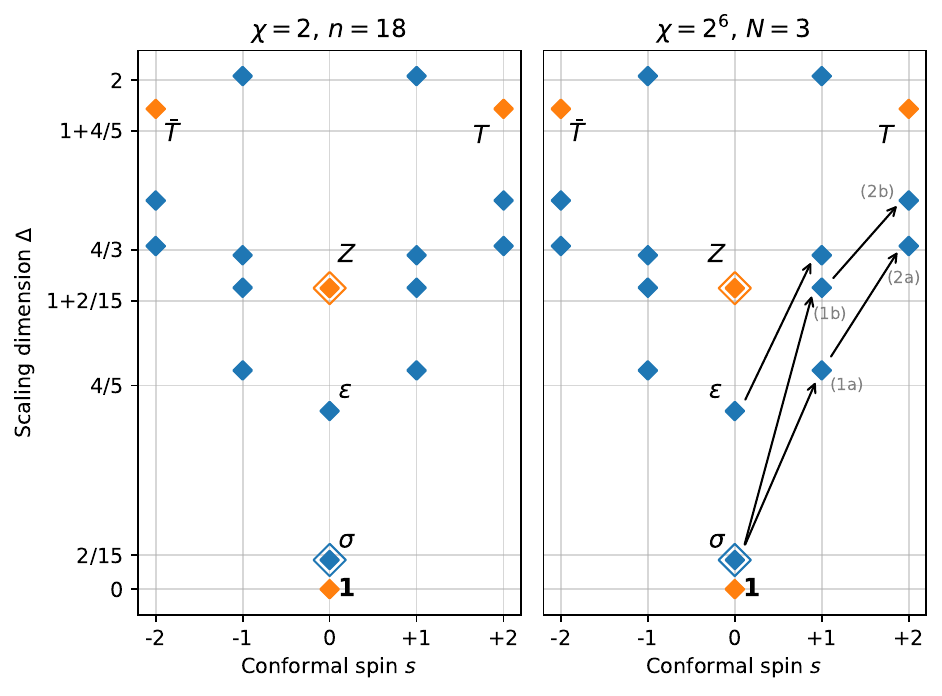}
  \caption{We use the cylinder of size $n=18$ to solve for the stress tensor (left). $T$ and $\bar{T}$ are used to built the Virasoro operators $L_n$ and $\bar{L}_n$ in a new cylinder of size $N=3$ and bond dimension $\chi=2^{n/3}$. The actions of $L_{-1}$ are illustrated by the arrows as an example (right). The states labeled (1a)/(1b) and (2a)/(2b) should be degenerate at the CFT limit but are split here due to the finite size effect (cf.\ \autoref{fig:fib-spectrum}). Note that to be consistent with the data in Appendix~\ref{sec:numerical-details}, the scaling dimensions shown here are \emph{not} rescaled.}
  \label{fig:fib-virasoro}
\end{figure}

In our calculation, we take cylinder size $n=12$, 15 and 18 to find the eigenstates corresponding to the stress tensor. They are reshaped and padded into 4-leg tensor units with bond dimension $\chi=2^{n/3}$. Then we use these tensor units as well as the blocking sites to form a new cylinder for obtaining the Virasoro operator. In principle, the size of the new cylinder $N$ can be chosen arbitrarily. But due to the limit of memory resource, we have to take an $N=3$ cylinder (equivalent to the $n$-site cylinder with bond dimension $\chi=2$) for the calculation of the stress tensor in reality. We check their validity by applying these operators to low-energy eigenstates (same from size $N$ cylinder) and see if they will be correctly raised or lowered into certain levels. We show the actions of $L_{-1}$ in \autoref{fig:fib-virasoro} and more numerical details can be found in Appendix~\ref{sec:numerical-details}.

\section{Conclusions}

In this paper, we inspect our proposal of implementing Virasoro symmetry in a tensor network in a more elaborate lattice model than the Ising model, Particularly, we worked with the Fibonacci model, which is known to suffer from more severe finite size effects. The method that we have proposed continues to work reasonably well --- although in this case, to extract the stress tensor from a relatively small transfer matrix tiling a small cylinder, we need to make use of the method of the idempotent, to project out states with different topological charges. Paired with the optimization procedure which fits the eigenvalues in a power law as it approaches the thermodynamic limit, we can isolate the appropriate ``seed'' state that would flow to the stress tensor. We show that using a seed state to construct Virasoro operator still produces reasonably accurate descendants in a cylinder that is marred by finite size effects. This complements our earlier work, giving further evidence with an improved algorithm for reconstructing Virasoro symmetry in a more general tensor network.

\mbox{}

\begin{acknowledgments}
  LYH acknowledges the support of NSFC (Grant No.\ 11922502, 11875111) and the Shanghai Municipal Science and Technology Major Project (Shanghai Grant No.\ 2019SHZDZX01), and Perimeter Institute for hospitality as a part of the Emmy Noether Fellowship Programme. We are particularly grateful to Prof.~Frank Verstraete and Robijn Vanhove for multiple correspondences, explaining in detail the methodologies they employed in their beautiful series of papers. We also thank Lin Chen, Jiaqi Lou, Bingxin Lao and Xinyang Yu for useful discussions and comments.
\end{acknowledgments}

\bibliography{main}

\onecolumngrid

\appendix

\section{Ising model}
\label{sec:ising-model}

Anyon types of the 2D critical Ising model on square lattice are $\1$, $\sigma$ and $\psi$, with quantum dimension $d_\1=d_\psi=1$, $d_\sigma=\sqrt2$. Non-trivial fusion rules and $F$-symbols are
\begin{gather}
  \sigma \times \sigma = \1 + \psi, \quad
  \sigma \times \psi   = \psi \times \sigma = \sigma, \quad
  \psi   \times \psi   = \1; \notag \\
  F^{\sigma\sigma\sigma}_\sigma = \frac{1}{\sqrt2} \begin{pmatrix} 1 & 1 \\ 1 & -1 \end{pmatrix}, \quad
  \bigl[ F^{\psi\sigma\psi}_\sigma \bigr]_{\sigma\sigma} = \bigl[ F^{\sigma\psi\sigma}_\psi \bigr]_{\sigma\sigma} = -1.
\end{gather}
We can then write the tetrahedra and triangle tensors as follows:
\begin{equation}
  \begin{aligned}
       \Triangle \sigma\sigma\1\1\sigma\sigma
    &= (d_\1 d_\sigma d_\sigma)  ^{-\frac14} (d_\1 d_\sigma d_\sigma)  ^{-\frac13}
       \Tetrahedron \1\sigma\sigma\1\sigma\sigma
       = 2^{-\frac{1}{12}}, \\
       \Triangle \sigma\sigma\1\psi\sigma\sigma
    &= (d_\1 d_\sigma d_\sigma)  ^{-\frac14} (d_\sigma d_\sigma d_\psi)^{-\frac13}
       \Tetrahedron \1\sigma\sigma\psi\sigma\sigma
       = 2^{-\frac{1}{12}}, \\
       \Triangle \sigma\sigma\psi\1\sigma\sigma
    &= (d_\sigma d_\sigma d_\psi)^{-\frac14} (d_\1 d_\sigma d_\sigma)  ^{-\frac13}
       \Tetrahedron \psi\sigma\sigma\1\sigma\sigma
       = 2^{-\frac{1}{12}}, \\
       \Triangle \sigma\sigma\psi\psi\sigma\sigma
    &= (d_\sigma d_\sigma d_\psi)^{-\frac14} (d_\sigma d_\sigma d_\psi)^{-\frac13}
       \Tetrahedron \psi\sigma\sigma\psi\sigma\sigma
       = -2^{-\frac{1}{12}}.
  \end{aligned}
\end{equation}
As for the strange correlator, we need to project physical legs to a direct product state
\begin{equation}
  \bra{\omega(\beta)} = \sqrt2 \bigl( \cosh\beta \bra{\1} + \sinh\beta \bra{\psi} \bigr).
\end{equation}
In order to follow the square lattice structure, we construct the tensor unit as an octagon containing 8 triangles:
\def\DiagramI{
  \begin{tikzpicture}[baseline=-0.5ex]
    \draw [tensor box]
          ( 0.3, 0.7) -- ( 1.9, 2.3) -- ( 0.3, 2.3) -- cycle
          ( 0.7, 0.3) -- ( 2.3, 0.3) -- ( 2.3, 1.9) -- cycle
          (-0.3, 0.7) -- (-1.9, 2.3) -- (-0.3, 2.3) -- cycle
          (-0.7, 0.3) -- (-2.3, 0.3) -- (-2.3, 1.9) -- cycle
          ( 0.3,-0.7) -- ( 1.9,-2.3) -- ( 0.3,-2.3) -- cycle
          ( 0.7,-0.3) -- ( 2.3,-0.3) -- ( 2.3,-1.9) -- cycle
          (-0.3,-0.7) -- (-1.9,-2.3) -- (-0.3,-2.3) -- cycle
          (-0.7,-0.3) -- (-2.3,-0.3) -- (-2.3,-1.9) -- cycle;
    % \sigma
    \draw [semithick, MaterialTeal]
          ( 1.1, 2.0) -- ( 1.1, 3.0) ( 2.0, 1.1) -- ( 3.0, 1.1)
          (-1.1, 2.0) -- (-1.1, 3.0) (-2.0, 1.1) -- (-3.0, 1.1)
          ( 1.1,-2.0) -- ( 1.1,-3.0) ( 2.0,-1.1) -- ( 3.0,-1.1)
          (-1.1,-2.0) -- (-1.1,-3.0) (-2.0,-1.1) -- (-3.0,-1.1)
          ( 0.6, 1.5) -- (-0.6, 1.5)
          ( 1.5, 0.6) -- ( 1.5,-0.6)
          ( 0.6,-1.5) -- (-0.6,-1.5)
          (-1.5, 0.6) -- (-1.5,-0.6)
          ( 1.0, 0.0) .. controls ( 1.0, 0.6) and ( 0.6, 1.0) .. ( 0.0, 1.0)
                      .. controls (-0.6, 1.0) and (-1.0, 0.6) .. (-1.0, 0.0)
                      .. controls (-1.0,-0.6) and (-0.6,-1.0) .. ( 0.0,-1.0)
                      .. controls ( 0.6,-1.0) and ( 1.0,-0.6) .. cycle
          ( 3.0, 1.5) .. controls ( 2.0, 1.5) and ( 1.5, 2.0) .. ( 1.5, 3.0)
          (-3.0, 1.5) .. controls (-2.0, 1.5) and (-1.5, 2.0) .. (-1.5, 3.0)
          ( 3.0,-1.5) .. controls ( 2.0,-1.5) and ( 1.5,-2.0) .. ( 1.5,-3.0)
          (-3.0,-1.5) .. controls (-2.0,-1.5) and (-1.5,-2.0) .. (-1.5,-3.0);
    % 1 \otimes \psi
    \draw [semithick, MaterialIndigo]
          ( 1.7, 0.9) -- ( 0.9, 1.7)
          (-1.7, 0.9) -- (-0.9, 1.7)
          ( 1.7,-0.9) -- ( 0.9,-1.7)
          (-1.7,-0.9) -- (-0.9,-1.7);
    % 1 or \psi
    \draw [tensor leg]
          ( 0.6, 3.0) .. controls ( 0.6, 2.6) and ( 0.6, 2.0) .. ( 0.0, 2.0)
                      .. controls (-0.6, 2.0) and (-0.6, 2.6) .. (-0.6, 3.0)
          ( 3.0, 0.6) .. controls ( 2.6, 0.6) and ( 2.0, 0.6) .. ( 2.0, 0.0)
                      .. controls ( 2.0,-0.6) and ( 2.6,-0.6) .. ( 3.0,-0.6)
          ( 0.6,-3.0) .. controls ( 0.6,-2.6) and ( 0.6,-2.0) .. ( 0.0,-2.0)
                      .. controls (-0.6,-2.0) and (-0.6,-2.6) .. (-0.6,-3.0)
          (-3.0, 0.6) .. controls (-2.6, 0.6) and (-2.0, 0.6) .. (-2.0, 0.0)
                      .. controls (-2.0,-0.6) and (-2.6,-0.6) .. (-3.0,-0.6);
    \draw ( 0.0, 3.5) node {$i$}
          (-3.5, 0.0) node {$j$}
          ( 0,.0-3.5) node {$k$}
          ( 3.5, 0.0) node {$l$};
  \end{tikzpicture}
}
\def\LineSigma{
  \begin{tikzpicture}[baseline=1.6ex]
    \draw [semithick, MaterialTeal] (0,0) -- (0,2);
  \end{tikzpicture}
}
\def\LineOmega{
  \begin{tikzpicture}[baseline=1.6ex]
    \draw [semithick, MaterialIndigo] (0,0) -- (0,2);
  \end{tikzpicture}
}
\begin{equation}
  A_{ijkl} = \DiagramI \quad \text{where} \quad
  i, j, k, l = \1 \text{ or } \psi, \quad
  \LineSigma \, = \sigma, \quad
  \LineOmega \, = \omega.
\end{equation}
To calculate the value of $A_{ijkl}$, we first check each pair of 2 triangles:
\def\PathI{
  \draw [tensor box]
        (-0.6,-1.0) -- ( 1.0,-1.0) -- ( 1.0, 0.6) -- cycle
        ( 0.6, 1.0) -- (-1.0, 1.0) -- (-1.0,-0.6) -- cycle;
  \draw [tensor leg]
        (-0.6, 1.7) .. controls (-0.6, 0.9) and (-0.9, 0.6) .. (-1.6, 0.6)
        ( 0.6,-1.7) .. controls ( 0.6,-0.9) and ( 0.9,-0.6) .. ( 1.6,-0.6);
  \draw [semithick, MaterialTeal]
        (-0.2, 0.7) -- (-0.2, 1.7) (-1.6, 0.2) -- (-0.6, 0.2)
        ( 0.2,-0.7) -- ( 0.2,-1.7) ( 1.6,-0.2) -- ( 0.6,-0.2)
        ( 1.6, 0.2) .. controls ( 0.7, 0.2) and ( 0.2, 0.7) .. ( 0.2, 1.7)
        (-1.6,-0.2) .. controls (-0.7,-0.2) and (-0.2,-0.7) .. (-0.2,-1.7);
  \draw [semithick, MaterialIndigo]
        ( 0.4,-0.4) -- (-0.4, 0.4);
  \draw (-1.5, 1.5) node {$i$}
        ( 1.5,-1.5) node {$j$};
}
\def\DiagramI{
  \begin{tikzpicture}[baseline=-0.5ex]
    \PathI
  \end{tikzpicture}
}
\def\DiagramII{
  \begin{tikzpicture}[baseline=-0.5ex, xscale=-1]
    \PathI
  \end{tikzpicture}
}
\begin{equation}
  \DiagramI = \DiagramII =
  \begin{cases}
    2^{\frac13} \bigl( \cosh\beta + \sinh\beta \bigr) = 2^{\frac13} \ee^\beta,
      & (i,j) = (\1,\1) \text{ or } (\psi, \psi); \\
    2^{\frac13} \bigl( \cosh\beta - \sinh\beta \bigr) = 2^{\frac13} \ee^{-\beta},
      & (i,j) = (\1,\psi) \text{ or } (\1, \psi).
  \end{cases}
\end{equation}
Then clearly
\begin{equation}
  A_{\1\1\1\1} = A_{\psi\psi\psi\psi} = 2^{\frac43} \ee^{4\beta}, \quad
  A_{\1\psi\1\psi} = A_{\psi\1\psi\1} = 2^{\frac43} \ee^{-4\beta}, \quad
  \text{others} = 2^{\frac43}.
  \label{eq:ising-tensor-from-sc}
\end{equation}
This is exactly the tensor representation of Ising interaction around a plaquette of spins:
\begin{equation}
  A_{ijkl}
    := A_{\sigma_i \sigma_j \sigma_k \sigma_l}
    = \ee^{\beta(\sigma_i\sigma_j + \sigma_j\sigma_k + \sigma_k\sigma_l + \sigma_l\sigma_i)},
  \label{eq:ising-tensor}
\end{equation}
up to some overall factor. Finally, we should mention that in this scheme, the critical temperature $\beta_{\text{c}}$ can be given via shifting the octagon by $1/2$ unit and employing the Kramers--Wannier duality.

\begin{figure}[ht]
  \centering
  \includegraphics[width=0.55\textwidth]{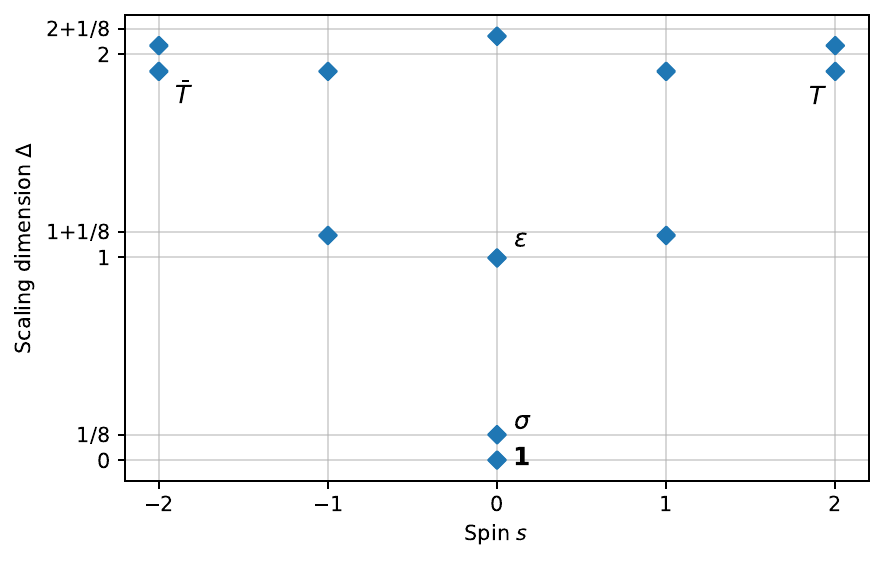}
  \caption{Ising spectrum corresponding to the cylinder transfer matrix of size $n\to\infty$. Primary states are labeled with their corresponding operators in the CFT.}
  \label{fig:ising-spectrum}
\end{figure}

Using the tensor network representation built from equations~\eqref{eq:ising-tensor-from-sc} and \eqref{eq:ising-tensor}, the partition function of 2D critical Ising model can now be written as
\begin{equation}
  Z = \sum_{\{\sigma\}} \prod_{\langle i,j \rangle} \ee^{\beta\sigma_i\sigma_j} = \lim_{m\to\infty} \Tr M^m,
\end{equation}
where $M$ is the transfer matrix:
\def\DiagramI{
  \begin{tikzpicture}[baseline=-0.5ex]
    \draw [tensor leg]
          (-1,1) .. controls (-2,1) and (-2,0) .. (-1,0) -- (5,0) (7,0) --
          ( 9,0) .. controls (10,0) and (10,1) .. ( 9,1);
    \draw [covered tensor leg] (-1,1) -- (9,1);
    \foreach \x/\xtext in {0/1, 2/2, 4/3, 8/n} {
      \draw [tensor leg] (\x,-1.5) -- (\x,1.5);
      \draw (\x,0) node [circle, draw, tensor box] {};
      \draw (\x    , 1.5) node [above] {$i_\xtext$}
            (\x    ,-1.5) node [below] {$k_\xtext$}
            (\x-0.8,-0.6) node {$j_\xtext$};
    }
    \draw (6,0) node [tensor leg] {...};
  \end{tikzpicture}
}
\begin{equation}
    M^{i_1 i_2 \cdots i_n}_{k_1 k_2 \cdots k_n}
  = \sum_{j_1, j_2, \ldots, j_n} \prod_{\alpha=1}^n A_{i_\alpha j_\alpha k_\alpha j_{\alpha+1}}
  = \DiagramI.
  \label{eq:cylinder-transfer-matrix}
\end{equation}
The Ising CFT spectrum can then be obtained by diagonalizing $T\cdot M$, where $T$ is the translation operator~\eqref{eq:translation-operator}. The spectrum data are listed in \autoref{tab:ising-spectrum}, where ``$\infty$'' column is extrapolated from the raw data with $\Delta=A+B/n$. We can visualize them in \autoref{fig:ising-spectrum}.

\begin{table}[ht]
  \centering\small
  \begin{tabular}{*{9}{c}}
    \toprule
      \header{Spin}{Size} & 4        & 8        & 12       & 16       & 20       & 24       & $\infty$     & Exact \\
    \midrule
      0                   & 0.123499 & 0.124606 & 0.124823 & 0.124900 & 0.124936 & 0.124955 & 0.12522\,(6) & $\frac18$     \\
      0                   & 1.026740 & 1.006488 & 1.002868 & 1.001610 & 1.001030 & 1.000715 & 0.9961\,(11) & 1             \\
      $\pm1$              & 1.245699 & 1.151345 & 1.136446 & 1.131388 & 1.129074 & 1.127824 & 1.107\,(5)   & $1{+}\frac18$ \\
      $\pm1,\pm2$         & 2.569513 & 2.098365 & 2.041544 & 2.022974 & 2.014590 & 2.010090 & 1.916\,(30)  & 2             \\
      0                   & 2.367899 & 2.178085 & 2.148069 & 2.137876 & 2.133212 & 2.130692 & 2.089\,(11)  & $2{+}\frac18$ \\
      $\pm2$              & -        & 2.369005 & 2.223018 & 2.178379 & 2.158670 & 2.148201 & 2.043\,(19)  & $2{+}\frac18$ \\
    \bottomrule
  \end{tabular}
  \caption{Ising spectrum. The cylinder size is from 4 to 24 and is extrapolated to infinity, with data listed in the ``$\infty$'' column.}
  \label{tab:ising-spectrum}
\end{table}

\section{Fitting details}
\label{sec:fitting-details}

\begin{figure}[ht]
  \centering
  \includegraphics[width=\textwidth]{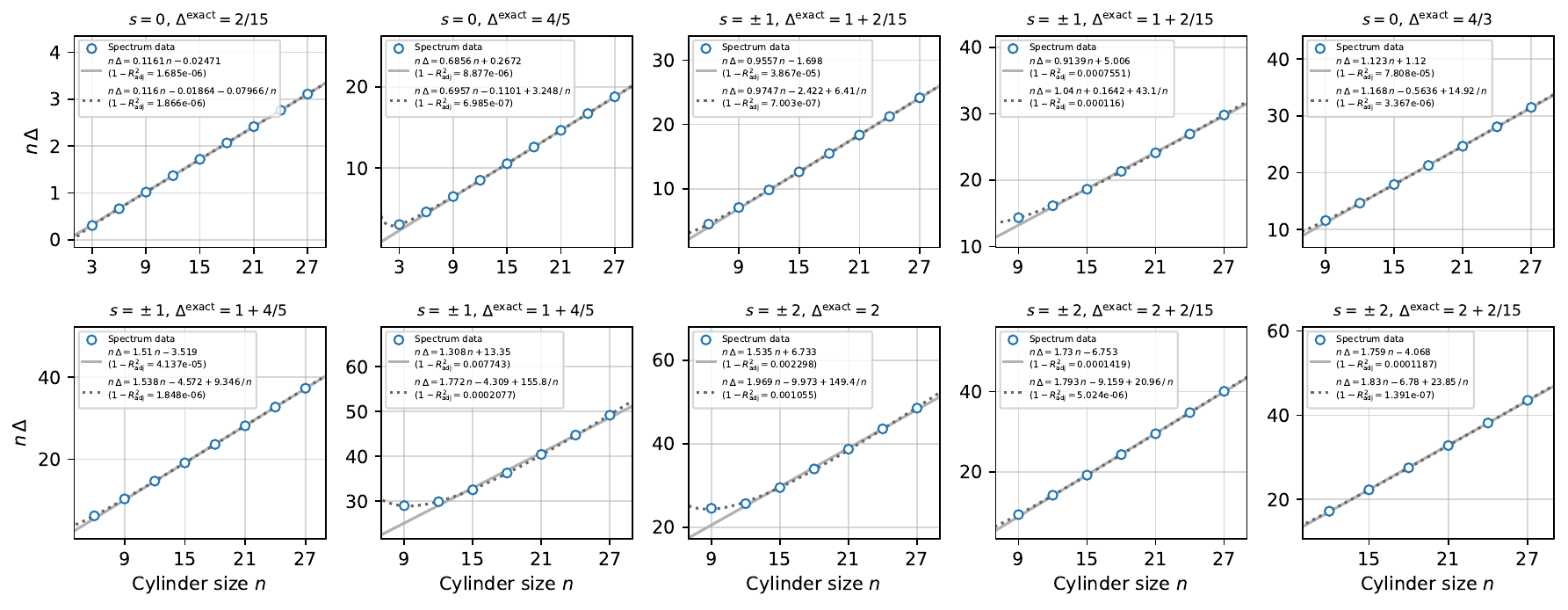}
  \caption{Fitting results for Fibonacci spectrum. Note that we only use cylinder sizes $n$ from 12 to 27 in fitting $n\Delta=An+B$ (solid lines), but all available $n$ in fitting $n\Delta=An+B+C/n$ (dotted lines).}
  \label{fig:fib-fitting}
\end{figure}

For a 2D finite size lattice model, the scaling dimensions $\Delta$ will approach the thermodynamic limiting values as $\Delta=A+B/n$, so $n\Delta$ - $n$ graph will be an approximate straight line. Hence $\Delta^\infty$ can be obtained by fitting these straight lines with small cylinders omitted (see the solid lines in \autoref{fig:fib-fitting}). They are listed in the ``$\infty$'' column in \autoref{tab:fib-spectrum}.

To further improve the accuracy, we can also fit the spectrum data with higher order corrections, such as $\Delta=A+B/n+C/n^2$ (dotted lines in \autoref{fig:fib-fitting}). This will give a better fit to the data, but as to identify the required states in the CFT, only $1/n$ correction is needed.

\section{Implementation details}
\label{sec:implementation-details}

\begin{figure}[ht]
  \input{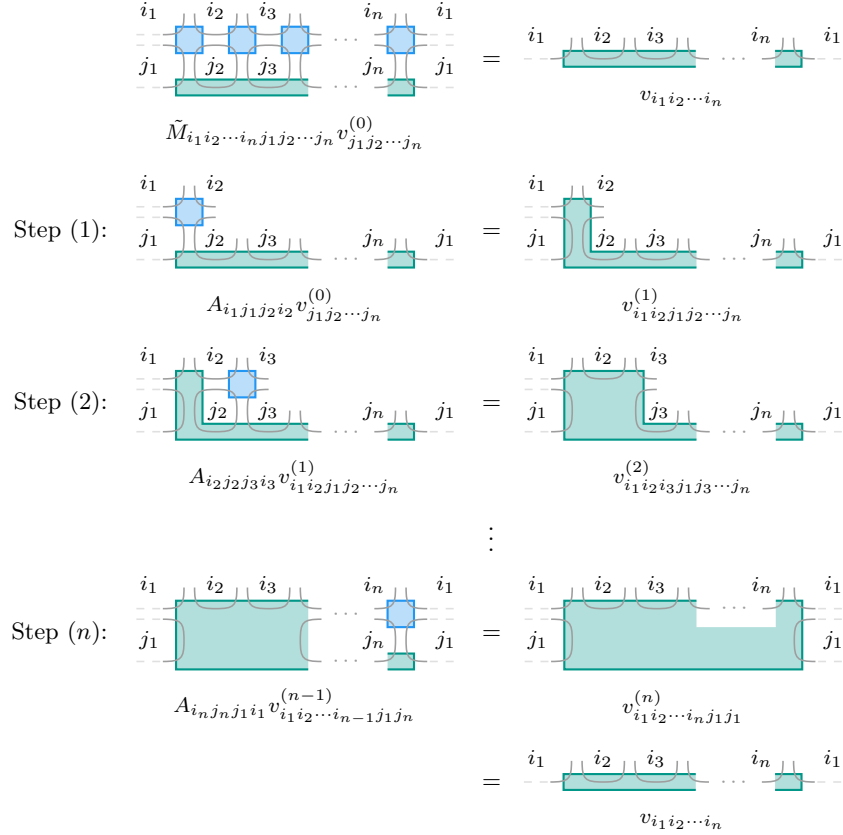}
  \caption{Implementation of the matrix-vector multiplication $M\cdot v$ in the LinearOperator method for Fibonacci string-net.}
  \label{fig:linear-operator}
\end{figure}

The implementation of the matrix-vector multiplication in the LinearOperator method is illustrated in \autoref{fig:linear-operator}. Since the transfer matrix $M$ has already been written in a ring of $A$ tensors, the multiplication $M\cdot v$ is then constructed by applying the tensor units iteratively. To obtain the translation operator equation~\eqref{eq:translation-operator}, we can permute the elements of $v$ by a simple array indexing operation. For the $n$-site cylinder with bond dimension $\chi$, LinearOperator method can largely reduce the memory consumption from $\mathcal{O}(\chi^{2n})$ to $\mathcal{O}(\chi^{n})$, which makes the calculation of large cylinders possible. We use the standard Arnoldi algorithm for the diagonalization of the transfer matrix, where the LinearOperator method can also apply.

\section{Numerical details for the Virasoro operators}
\label{sec:numerical-details}

We check the validity of the lattice Virasoro operators (obtained from the cylinder eigenstates of size $n=18$) by applying them to low-energy eigenstates of a coarse-grained lattice with cylinder size $N=3$. The matrix elements $\mel{\phi_\alpha}{L_n}{\phi_\beta}$ for $n=\pm1$ and $\pm2$ are listed in \autoref{tab:L-m1}--\ref{tab:L-p2} and their behaviors on the conformal towers are illustrated in \autoref{fig:fib-virasoro-1} and \ref{fig:fib-virasoro-2}. In the tables, the matrix elements between states and their correct Virasoro descendants are highlighted in bold blue, whereas the erroneous matrix elements are marked in pale blue.

\begingroup

\catcode`\"=\active
\catcode`\'=\active
\def"#1"{\textcolor{MaterialBlue700}{\bfseries#1}}
\def'#1'{\textcolor{MaterialBlue300}{#1}}

\begin{figure}[ht]
  \includegraphics[width=\textwidth]{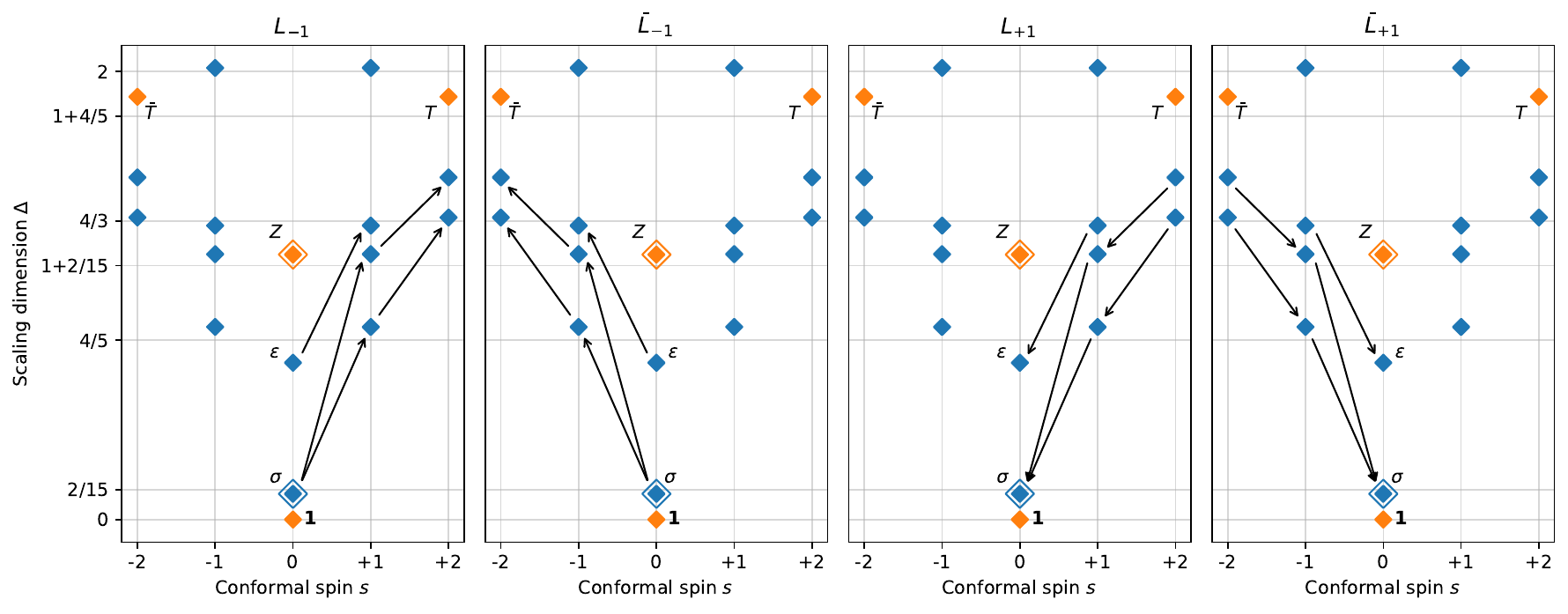}
  \caption{Actions of lattice $L_{\pm1}$ and $\bar{L}_{\pm1}$ in the low-energy subspace of cylinder transfer matrix of size $N=3$ and bond dimension $\chi=64$. Corresponding matrix elements are listed in \autoref{tab:L-m1}--\ref{tab:L-bar-p1}.}
  \label{fig:fib-virasoro-1}
\end{figure}

\begin{table}[htb]
  \centering\small
  \def\M#1{$\mel{\phi_\alpha}{L_{-1}}{#1}$}
  \begin{tabular}{cccc*{7}{c}}
    \toprule
      $\ket{\phi_\alpha}$ & $\Delta_\alpha$ & $\Delta^{\text{exact}}_\alpha$ & $s_\alpha$ &
      \M2 & \M3 & \M4 & \M5 & \M6 & \M7 & \M8 \\
    \midrule
      1  & 0.000 & 0                  & 0    &  1.668e-16  &  1.063e-16  &  2.449e-16  &  2.247e-16  &  1.924e-02  &  1.837e-16  &  1.297e-02  \\
      2  & 0.115 & $    \frac{2}{15}$ & 0    &  1.332e-16  &  1.264e-16  &  6.993e-17  &  1.013e-15  &  3.914e-02  &  1.133e-16  & '1.395e-01' \\
      3  & 0.115 & $    \frac{2}{15}$ & 0    &  2.676e-16  &  8.564e-17  &  9.721e-17  &  2.311e-16  & '1.436e-01' &  3.369e-16  &  3.779e-02  \\
      4  & 0.700 & $    \frac{4}{5} $ & 0    &  9.051e-16  &  9.529e-16  &  2.877e-16  &  8.961e-16  & '1.528e-01' &  4.405e-16  & '1.456e-01' \\
      5  & 1.185 & $1{+}\frac{2}{15}$ & $+1$ & "9.653e-01" & '9.059e-02' &  2.550e-02  &  4.998e-16  &  1.203e-16  &  1.033e-16  &  1.854e-16  \\
      6  & 0.860 & $1{+}\frac{2}{15}$ & $-1$ &  3.743e-15  &  1.485e-14  &  1.905e-15  &  3.670e-02  &  9.813e-16  & '5.692e-01' &  4.330e-16  \\
      7  & 0.860 & $1{+}\frac{2}{15}$ & $+1$ & '1.054e-01' & "9.112e-01" &  5.579e-02  &  6.870e-17  &  1.497e-16  &  2.763e-16  &  6.323e-16  \\
      8  & 1.185 & $1{+}\frac{2}{15}$ & $-1$ &  4.086e-14  &  5.837e-15  &  1.360e-15  & '8.688e-01' &  3.384e-16  &  1.440e-02  &  4.357e-15  \\
      9  & 1.183 & $    \frac{4}{3} $ & 0    &  1.408e-14  &  2.396e-15  &  1.321e-15  &  9.665e-16  &  6.665e-03  &  5.264e-16  &  2.095e-02  \\
      10 & 1.183 & $    \frac{4}{3} $ & 0    &  9.490e-15  &  2.714e-15  &  3.248e-16  &  1.626e-15  &  2.858e-03  &  2.244e-15  &  2.560e-02  \\
      11 & 1.312 & $1{+}\frac{4}{5} $ & $+1$ & '9.616e-02' & '2.486e-01' & "8.767e-01" &  3.467e-16  &  2.227e-16  &  7.391e-17  &  2.155e-16  \\
      12 & 1.312 & $1{+}\frac{4}{5} $ & $-1$ &  9.042e-16  &  3.220e-15  &  1.149e-14  & '2.295e-01' &  3.092e-16  & '2.371e-01' &  1.272e-15  \\
      13 & 1.887 & 2                  & $+2$ &  4.287e-16  &  1.287e-15  &  4.530e-15  &  9.052e-03  &  7.307e-16  &  7.514e-02  &  2.232e-16  \\
      14 & 1.887 & 2                  & $-2$ &  5.546e-02  &  3.241e-02  & '2.098e-01' &  1.801e-15  &  5.093e-16  &  1.742e-15  &  2.569e-15  \\
      15 & 1.527 & $2{+}\frac{2}{15}$ & $+2$ &  1.466e-15  &  1.605e-16  &  6.354e-16  & "2.863e-01" &  6.779e-16  & '1.974e-01' &  1.335e-15  \\
      16 & 1.348 & $2{+}\frac{2}{15}$ & $-2$ &  1.480e-03  & '6.605e-01' & '3.804e-01' &  1.088e-15  &  1.692e-16  &  4.408e-15  &  4.819e-16  \\
      17 & 1.348 & $2{+}\frac{2}{15}$ & $+2$ &  3.851e-16  &  5.451e-16  &  2.807e-16  & '9.368e-02' &  1.927e-15  & "9.331e-01" &  9.124e-16  \\
      18 & 1.527 & $2{+}\frac{2}{15}$ & $-2$ &  7.978e-03  &  2.011e-02  & '7.351e-01' &  3.228e-15  &  2.967e-16  &  1.856e-15  &  6.081e-16  \\
    \bottomrule
  \end{tabular}
  \caption{\M{\phi_\beta}}
  \label{tab:L-m1}
\end{table}

\begin{table}[htb]
  \centering\small
  \def\M#1{$\mel{\phi_\alpha}{\bar{L}_{-1}}{#1}$}
  \begin{tabular}{cccc*{7}{c}}
    \toprule
      $\ket{\phi_\alpha}$ & $\Delta_\alpha$ & $\Delta^{\text{exact}}_\alpha$ & $s_\alpha$ &
      \M2 & \M3 & \M4 & \M5 & \M6 & \M7 & \M8 \\
    \midrule
      1  & 0.000 & 0                  & 0    &  1.332e-16  &  1.782e-16  &  2.228e-16  &  1.297e-02  &  1.755e-16  &  1.924e-02  &  1.980e-16  \\
      2  & 0.115 & $    \frac{2}{15}$ & 0    &  6.331e-17  &  2.548e-16  &  8.701e-17  &  3.779e-02  &  3.451e-16  & '1.436e-01' &  2.382e-16  \\
      3  & 0.115 & $    \frac{2}{15}$ & 0    &  1.351e-16  &  1.343e-16  &  6.665e-17  & '1.395e-01' &  1.049e-16  &  3.914e-02  &  1.013e-15  \\
      4  & 0.700 & $    \frac{4}{5} $ & 0    &  9.563e-16  &  8.944e-16  &  2.835e-16  & '1.456e-01' &  4.352e-16  & '1.528e-01' &  8.986e-16  \\
      5  & 1.185 & $1{+}\frac{2}{15}$ & $+1$ &  5.836e-15  &  4.088e-14  &  1.342e-15  &  4.295e-15  &  1.440e-02  &  2.886e-16  & '8.688e-01' \\
      6  & 0.860 & $1{+}\frac{2}{15}$ & $-1$ & "9.112e-01" & '1.054e-01' &  5.579e-02  &  6.370e-16  &  2.610e-16  &  1.433e-16  &  8.604e-17  \\
      7  & 0.860 & $1{+}\frac{2}{15}$ & $+1$ &  1.489e-14  &  3.625e-15  &  1.958e-15  &  4.199e-16  & '5.692e-01' &  8.834e-16  &  3.670e-02  \\
      8  & 1.185 & $1{+}\frac{2}{15}$ & $-1$ & '9.059e-02' & "9.653e-01" &  2.550e-02  &  1.756e-16  &  1.109e-16  &  1.284e-16  &  5.018e-16  \\
      9  & 1.183 & $    \frac{4}{3} $ & 0    &  2.780e-15  &  9.422e-15  &  3.590e-16  &  2.560e-02  &  2.174e-15  &  2.858e-03  &  1.636e-15  \\
      10 & 1.183 & $    \frac{4}{3} $ & 0    &  2.363e-15  &  1.403e-14  &  1.313e-15  &  2.095e-02  &  5.253e-16  &  6.665e-03  &  9.721e-16  \\
      11 & 1.312 & $1{+}\frac{4}{5} $ & $+1$ &  3.269e-15  &  8.494e-16  &  1.152e-14  &  1.212e-15  & '2.371e-01' &  3.656e-16  & '2.295e-01' \\
      12 & 1.312 & $1{+}\frac{4}{5} $ & $-1$ & '2.486e-01' & '9.616e-02' & "8.767e-01" &  1.973e-16  &  7.044e-17  &  2.180e-16  &  3.385e-16  \\
      13 & 1.887 & 2                  & $+2$ &  3.241e-02  &  5.546e-02  & '2.098e-01' &  2.547e-15  &  1.751e-15  &  5.323e-16  &  1.771e-15  \\
      14 & 1.887 & 2                  & $-2$ &  1.216e-15  &  4.646e-16  &  4.497e-15  &  2.205e-16  &  7.514e-02  &  7.693e-16  &  9.052e-03  \\
      15 & 1.527 & $2{+}\frac{2}{15}$ & $+2$ &  2.011e-02  &  7.978e-03  & '7.351e-01' &  6.155e-16  &  1.839e-15  &  2.672e-16  &  3.215e-15  \\
      16 & 1.348 & $2{+}\frac{2}{15}$ & $-2$ &  5.333e-16  &  3.697e-16  &  2.794e-16  &  9.403e-16  & "9.331e-01" &  1.924e-15  & '9.368e-02' \\
      17 & 1.348 & $2{+}\frac{2}{15}$ & $+2$ & '6.605e-01' &  1.480e-03  & '3.804e-01' &  4.580e-16  &  4.341e-15  &  2.112e-16  &  1.088e-15  \\
      18 & 1.527 & $2{+}\frac{2}{15}$ & $-2$ &  1.368e-16  &  1.532e-15  &  6.654e-16  &  1.282e-15  & '1.974e-01' &  6.726e-16  & "2.863e-01" \\
    \bottomrule
  \end{tabular}
  \caption{\M{\phi_\beta}}
  \label{tab:L-bar-m1}
\end{table}

\begin{table}[htb]
  \centering\small
  \def\M#1{$\mel{\phi_\alpha}{L_{+1}}{#1}$}
  \begin{tabular}{cccc*{9}{c}}
    \toprule
      $\ket{\phi_\alpha}$ & $\Delta_\alpha$ & $\Delta^{\text{exact}}_\alpha$ & $s_\alpha$ &
      \M5 & \M6 & \M7 & \M8 & \M{11} & \M{12} & \M{15} & \M{17} \\
    \midrule
      1  & 0.000 & 0                  & 0    & '1.039e-01' &  2.787e-16  & '1.357e-01' &  5.910e-16  &  1.975e-02  &  6.362e-17  &  1.446e-16  &  7.152e-17  \\
      2  & 0.115 & $    \frac{2}{15}$ & 0    & "9.009e-01" &  5.282e-16  &  7.788e-02  &  4.695e-15  &  2.377e-02  &  1.748e-16  &  3.096e-16  &  2.613e-16  \\
      3  & 0.115 & $    \frac{2}{15}$ & 0    & '1.076e-01' &  1.699e-15  & "9.004e-01" &  8.795e-16  &  1.228e-02  &  1.804e-16  &  1.009e-16  &  2.888e-16  \\
      4  & 0.700 & $    \frac{4}{5} $ & 0    &  3.127e-02  &  9.127e-17  &  3.978e-02  &  4.922e-16  & "6.543e-01" &  1.632e-15  &  3.342e-16  &  3.956e-16  \\
      5  & 1.185 & $1{+}\frac{2}{15}$ & $+1$ &  1.093e-15  &  1.126e-02  &  3.602e-16  & '2.028e-01' &  4.985e-16  & '1.280e-01' & "2.043e-01" &  7.993e-02  \\
      6  & 0.860 & $1{+}\frac{2}{15}$ & $-1$ &  2.469e-16  &  7.992e-16  &  2.700e-16  &  3.027e-16  &  1.075e-16  &  3.936e-16  &  4.661e-16  &  5.142e-15  \\
      7  & 0.860 & $1{+}\frac{2}{15}$ & $+1$ &  4.036e-17  & '1.342e-01' &  4.806e-16  &  3.159e-03  &  1.523e-16  &  1.822e-02  &  1.384e-02  & "8.273e-01" \\
      8  & 1.185 & $1{+}\frac{2}{15}$ & $-1$ &  5.780e-16  &  2.047e-16  &  2.252e-15  &  2.400e-15  &  5.186e-16  &  1.352e-15  &  2.230e-15  &  1.923e-15  \\
      9  & 1.183 & $    \frac{4}{3} $ & 0    &  1.991e-02  &  2.422e-16  &  5.860e-04  &  9.291e-16  &  6.752e-02  &  5.515e-16  &  8.351e-16  &  5.743e-16  \\
      10 & 1.183 & $    \frac{4}{3} $ & 0    &  4.026e-02  &  8.613e-17  &  1.138e-02  &  5.202e-16  & '9.652e-02' &  2.857e-16  &  7.076e-16  &  9.802e-16  \\
      11 & 1.312 & $1{+}\frac{4}{5} $ & $+1$ &  8.062e-16  & '8.189e-02' &  4.240e-16  & '1.411e-01' &  2.285e-16  &  2.620e-02  &  5.288e-02  & '8.538e-02' \\
      12 & 1.312 & $1{+}\frac{4}{5} $ & $-1$ &  8.267e-17  &  2.347e-16  &  4.970e-16  &  3.741e-17  &  4.512e-16  &  1.381e-16  &  2.376e-16  &  5.710e-16  \\
      13 & 1.887 & 2                  & $+2$ &  2.734e-16  &  2.423e-16  &  5.158e-17  &  1.735e-16  &  1.009e-16  &  2.237e-16  &  1.145e-16  &  8.372e-17  \\
      14 & 1.887 & 2                  & $-2$ &  1.707e-15  & '1.140e-01' &  1.772e-15  &  5.354e-02  &  1.110e-15  &  5.932e-02  &  6.955e-02  & '8.001e-02' \\
      15 & 1.527 & $2{+}\frac{2}{15}$ & $+2$ &  9.009e-16  &  1.473e-16  &  8.538e-17  &  2.064e-16  &  4.773e-17  &  1.280e-16  &  6.262e-17  &  9.921e-17  \\
      16 & 1.348 & $2{+}\frac{2}{15}$ & $-2$ &  1.737e-15  & '1.919e-01' &  9.190e-16  &  3.725e-03  &  4.467e-16  & '1.090e-01' & '2.726e-01' & '4.972e-01' \\
      17 & 1.348 & $2{+}\frac{2}{15}$ & $+2$ &  4.061e-17  &  1.464e-16  &  3.346e-16  &  4.723e-16  &  1.234e-16  &  1.344e-16  &  1.100e-16  &  4.950e-16  \\
      18 & 1.527 & $2{+}\frac{2}{15}$ & $-2$ &  3.148e-16  & '5.491e-01' &  1.778e-15  &  6.810e-02  &  2.825e-16  & '9.379e-02' &  1.075e-02  & '3.427e-01' \\
    \bottomrule
  \end{tabular}
  \caption{\M{\phi_\beta}}
  \label{tab:L-p1}
\end{table}

\begin{table}[htb]
  \centering\small
  \def\M#1{$\mel{\phi_\alpha}{\bar{L}_{+1}}{#1}$}
  \begin{tabular}{cccc*{9}{c}}
    \toprule
      $\ket{\phi_\alpha}$ & $\Delta_\alpha$ & $\Delta^{\text{exact}}_\alpha$ & $s_\alpha$ &
      \M5 & \M6 & \M7 & \M8 & \M{11} & \M{12} & \M{16} & \M{18} \\
    \midrule
      1  & 0.000 & 0                  & 0    &  5.855e-16  & '1.357e-01' &  2.693e-16  & '1.039e-01' &  8.166e-17  &  1.975e-02  &  1.041e-16  &  1.189e-16  \\
      2  & 0.115 & $    \frac{2}{15}$ & 0    &  9.024e-16  & "9.004e-01" &  1.656e-15  & '1.076e-01' &  1.928e-16  &  1.228e-02  &  2.915e-16  &  1.348e-16  \\
      3  & 0.115 & $    \frac{2}{15}$ & 0    &  4.735e-15  &  7.788e-02  &  5.232e-16  & "9.009e-01" &  1.782e-16  &  2.377e-02  &  2.380e-16  &  3.062e-16  \\
      4  & 0.700 & $    \frac{4}{5} $ & 0    &  5.104e-16  &  3.978e-02  &  1.032e-16  &  3.127e-02  &  1.610e-15  & "6.543e-01" &  3.864e-16  &  3.288e-16  \\
      5  & 1.185 & $1{+}\frac{2}{15}$ & $+1$ &  2.385e-15  &  2.248e-15  &  2.051e-16  &  5.693e-16  &  1.353e-15  &  5.051e-16  &  1.939e-15  &  2.248e-15  \\
      6  & 0.860 & $1{+}\frac{2}{15}$ & $-1$ &  3.159e-03  &  4.519e-16  & '1.342e-01' &  4.389e-17  &  1.822e-02  &  1.569e-16  & "8.273e-01" &  1.384e-02  \\
      7  & 0.860 & $1{+}\frac{2}{15}$ & $+1$ &  3.099e-16  &  2.725e-16  &  8.148e-16  &  2.299e-16  &  3.277e-16  &  1.078e-16  &  5.155e-15  &  4.757e-16  \\
      8  & 1.185 & $1{+}\frac{2}{15}$ & $-1$ & '2.028e-01' &  3.572e-16  &  1.126e-02  &  1.066e-15  & '1.280e-01' &  4.561e-16  &  7.993e-02  & "2.043e-01" \\
      9  & 1.183 & $    \frac{4}{3} $ & 0    &  4.994e-16  &  1.138e-02  &  6.255e-17  &  4.026e-02  &  2.829e-16  & '9.652e-02' &  9.567e-16  &  6.775e-16  \\
      10 & 1.183 & $    \frac{4}{3} $ & 0    &  9.015e-16  &  5.860e-04  &  2.327e-16  &  1.991e-02  &  5.194e-16  &  6.752e-02  &  5.520e-16  &  8.260e-16  \\
      11 & 1.312 & $1{+}\frac{4}{5} $ & $+1$ &  4.673e-17  &  4.944e-16  &  2.249e-16  &  8.800e-17  &  1.418e-16  &  4.762e-16  &  5.909e-16  &  2.370e-16  \\
      12 & 1.312 & $1{+}\frac{4}{5} $ & $-1$ & '1.411e-01' &  4.286e-16  & '8.189e-02' &  8.016e-16  &  2.620e-02  &  2.180e-16  & '8.538e-02' &  5.288e-02  \\
      13 & 1.887 & 2                  & $+2$ &  5.354e-02  &  1.809e-15  & '1.140e-01' &  1.732e-15  &  5.932e-02  &  1.075e-15  & '8.001e-02' &  6.955e-02  \\
      14 & 1.887 & 2                  & $-2$ &  1.717e-16  &  7.352e-17  &  2.281e-16  &  3.081e-16  &  2.331e-16  &  1.036e-16  &  7.397e-17  &  1.289e-16  \\
      15 & 1.527 & $2{+}\frac{2}{15}$ & $+2$ &  6.810e-02  &  1.882e-15  & '5.491e-01' &  3.442e-16  & '9.379e-02' &  3.299e-16  & '3.427e-01' &  1.075e-02  \\
      16 & 1.348 & $2{+}\frac{2}{15}$ & $-2$ &  4.734e-16  &  3.378e-16  &  1.229e-16  &  1.980e-17  &  1.401e-16  &  1.297e-16  &  5.183e-16  &  1.179e-16  \\
      17 & 1.348 & $2{+}\frac{2}{15}$ & $+2$ &  3.725e-03  &  9.342e-16  & '1.919e-01' &  1.768e-15  & '1.090e-01' &  4.343e-16  & '4.972e-01' & '2.726e-01' \\
      18 & 1.527 & $2{+}\frac{2}{15}$ & $-2$ &  1.846e-16  &  7.373e-17  &  1.417e-16  &  8.890e-16  &  1.367e-16  &  2.964e-17  &  8.957e-17  &  6.303e-17  \\
    \bottomrule
  \end{tabular}
  \caption{\M{\phi_\beta}}
  \label{tab:L-bar-p1}
\end{table}

\begin{figure}[ht]
  \includegraphics[width=\textwidth]{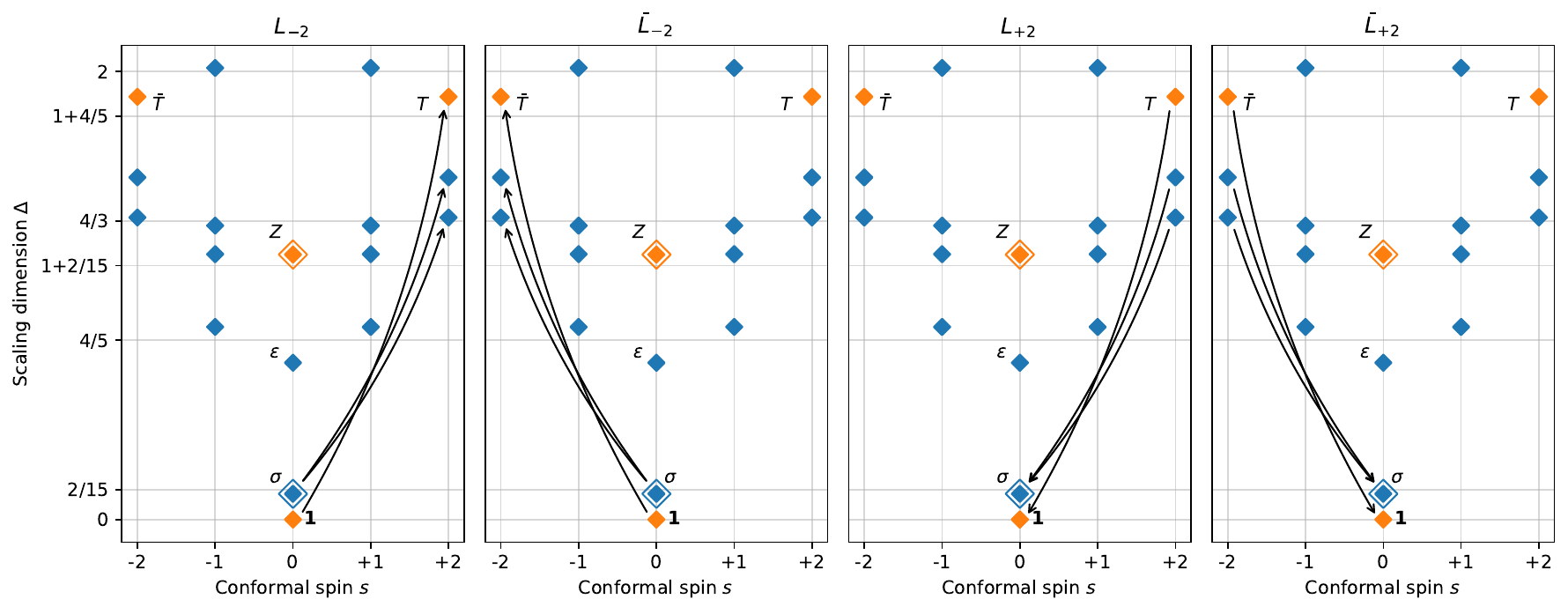}
  \caption{Actions of lattice $L_{\pm2}$ and $\bar{L}_{\pm2}$ in the low-energy subspace of cylinder transfer matrix of size $N=3$ and bond dimension $\chi=64$. Corresponding matrix elements are listed in \autoref{tab:L-m2} and \ref{tab:L-p2}.}
  \label{fig:fib-virasoro-2}
\end{figure}

\begin{table}[htb]
  \centering\small
  \def\M#1{$\mel{\phi_\alpha}{L_{-2}}{#1}$}
  \def\MB#1{$\mel{\phi_\alpha}{\bar{L}_{-2}}{#1}$}
  \begin{tabular}{cccc*{6}{c}}
    \toprule
      $\ket{\phi_\alpha}$ & $\Delta_\alpha$ & $\Delta^{\text{exact}}_\alpha$ & $s_\alpha$ &
      \M1 & \M2 & \M3 & \MB1 & \MB2 & \MB3 \\
    \midrule
      1  & 0.000 & 0                  & 0    &  5.070e-16  &  4.120e-16  &  5.410e-17  &  4.914e-16  &  4.335e-17  &  3.879e-16  \\
      2  & 0.115 & $    \frac{2}{15}$ & 0    &  2.123e-16  &  2.762e-16  &  1.842e-16  &  1.521e-16  &  9.780e-17  &  1.980e-16  \\
      3  & 0.115 & $    \frac{2}{15}$ & 0    &  9.160e-17  &  1.961e-16  &  7.979e-17  &  2.055e-16  &  1.908e-16  &  2.668e-16  \\
      4  & 0.700 & $    \frac{4}{5} $ & 0    &  1.860e-16  &  2.123e-16  &  3.320e-16  &  1.826e-16  &  3.088e-16  &  2.244e-16  \\
      5  & 1.185 & $1{+}\frac{2}{15}$ & $+1$ &  4.027e-16  &  1.686e-15  &  6.714e-16  & '8.983e-02' & '1.722e-01' & '8.496e-01' \\
      6  & 0.860 & $1{+}\frac{2}{15}$ & $-1$ &  7.056e-02  & '1.513e-01' & '5.550e-01' &  7.152e-17  &  5.173e-16  &  1.546e-16  \\
      7  & 0.860 & $1{+}\frac{2}{15}$ & $+1$ &  9.127e-17  &  1.621e-16  &  5.064e-16  &  7.056e-02  & '5.550e-01' & '1.513e-01' \\
      8  & 1.185 & $1{+}\frac{2}{15}$ & $-1$ & '8.983e-02' & '8.496e-01' & '1.722e-01' &  4.163e-16  &  6.801e-16  &  1.651e-15  \\
      9  & 1.183 & $    \frac{4}{3} $ & 0    &  3.034e-16  &  1.516e-15  &  4.880e-16  &  1.413e-16  &  1.837e-15  &  1.510e-15  \\
      10 & 1.183 & $    \frac{4}{3} $ & 0    &  1.355e-16  &  1.478e-15  &  1.836e-15  &  3.118e-16  &  5.225e-16  &  1.521e-15  \\
      11 & 1.312 & $1{+}\frac{4}{5} $ & $+1$ &  8.784e-17  &  9.239e-16  &  3.474e-16  & '1.213e-01' & '1.645e-01' & '5.981e-01' \\
      12 & 1.312 & $1{+}\frac{4}{5} $ & $-1$ & '1.213e-01' & '5.981e-01' & '1.645e-01' &  1.110e-16  &  3.210e-16  &  9.200e-16  \\
      13 & 1.887 & 2                  & $+2$ & "9.866e-01" &  6.101e-02  &  6.156e-02  &  5.115e-14  &  6.659e-15  &  5.855e-15  \\
      14 & 1.887 & 2                  & $-2$ &  5.103e-14  &  5.967e-15  &  6.529e-15  & "9.866e-01" &  6.156e-02  &  6.101e-02  \\
      15 & 1.527 & $2{+}\frac{2}{15}$ & $+2$ & '1.242e-01' & "9.650e-01" &  8.627e-03  &  7.442e-15  &  1.471e-15  &  5.853e-14  \\
      16 & 1.348 & $2{+}\frac{2}{15}$ & $-2$ &  1.741e-15  &  2.444e-15  &  7.129e-15  &  5.512e-02  & "9.282e-01" & '1.594e-01' \\
      17 & 1.348 & $2{+}\frac{2}{15}$ & $+2$ &  5.512e-02  & '1.594e-01' & "9.282e-01" &  1.791e-15  &  7.104e-15  &  2.544e-15  \\
      18 & 1.527 & $2{+}\frac{2}{15}$ & $-2$ &  7.514e-15  &  5.851e-14  &  1.378e-15  & '1.242e-01' &  8.627e-03  & "9.650e-01" \\
    \bottomrule
  \end{tabular}
  \caption{\M{\phi_\beta} and \MB{\phi_\beta}}
  \label{tab:L-m2}
\end{table}

\begin{table}[htb]
  \centering\small
  \setlength{\tabcolsep}{1.5pt}
  \def\M#1{$\mel{\phi_\alpha}{L_{+2}}{#1}$}
  \def\MB#1{$\mel{\phi_\alpha}{\bar{L}_{+2}}{#1}$}
  \begin{tabular}{cccc*{8}{c}}
    \toprule
      $\ket{\phi_\alpha}$ & $\Delta_\alpha$ & $\Delta^{\text{exact}}_\alpha$ & $s_\alpha$ &
      \M{13} & \M{14} & \M{15} & \M{17} & \MB{13} & \MB{14} & \MB{16} & \MB{18}\\
    \midrule
      1  & 0.000 & 0                  & 0    & "8.968e-01" &  1.644e-15  &  7.211e-02  &  4.644e-02  &  1.590e-15  & "8.968e-01" &  4.644e-02  &  7.211e-02  \\
      2  & 0.115 & $    \frac{2}{15}$ & 0    &  6.005e-02  &  4.075e-16  & "7.117e-01" & '9.349e-02' &  1.876e-16  &  3.489e-02  & "7.512e-01" &  2.103e-02  \\
      3  & 0.115 & $    \frac{2}{15}$ & 0    &  3.489e-02  &  1.714e-16  &  2.103e-02  & "7.512e-01" &  3.855e-16  &  6.005e-02  & '9.349e-02' & "7.117e-01" \\
      4  & 0.700 & $    \frac{4}{5} $ & 0    &  4.257e-03  &  1.786e-16  &  3.160e-02  &  2.784e-02  &  1.794e-16  &  4.257e-03  &  2.784e-02  &  3.160e-02  \\
      5  & 1.185 & $1{+}\frac{2}{15}$ & $+1$ &  1.357e-16  &  2.602e-16  &  2.121e-16  &  5.904e-17  &  2.340e-02  &  4.367e-16  &  2.207e-15  &  4.641e-16  \\
      6  & 0.860 & $1{+}\frac{2}{15}$ & $-1$ &  3.538e-16  &  2.998e-02  &  3.029e-16  &  3.710e-16  &  1.428e-16  &  4.351e-17  &  3.006e-16  &  1.212e-16  \\
      7  & 0.860 & $1{+}\frac{2}{15}$ & $+1$ &  3.852e-17  &  1.425e-16  &  1.114e-16  &  3.006e-16  &  2.998e-02  &  3.067e-16  &  4.008e-16  &  3.221e-16  \\
      8  & 1.185 & $1{+}\frac{2}{15}$ & $-1$ &  3.408e-16  &  2.340e-02  &  4.807e-16  &  2.198e-15  &  2.494e-16  &  1.399e-16  &  6.533e-17  &  2.065e-16  \\
      9  & 1.183 & $    \frac{4}{3} $ & 0    & '9.324e-02' &  1.388e-16  &  2.006e-02  &  3.345e-02  &  2.233e-16  &  7.031e-02  &  1.587e-02  &  5.714e-02  \\
      10 & 1.183 & $    \frac{4}{3} $ & 0    &  7.031e-02  &  2.239e-16  &  5.714e-02  &  1.587e-02  &  1.457e-16  & '9.324e-02' &  3.345e-02  &  2.006e-02  \\
      11 & 1.312 & $1{+}\frac{4}{5} $ & $+1$ &  1.082e-16  &  2.425e-16  &  1.205e-16  &  5.608e-17  &  5.229e-02  &  4.634e-16  &  3.855e-16  &  7.252e-16  \\
      12 & 1.312 & $1{+}\frac{4}{5} $ & $-1$ &  4.306e-16  &  5.229e-02  &  6.871e-16  &  3.659e-16  &  2.526e-16  &  1.292e-16  &  7.423e-17  &  1.092e-16  \\
      13 & 1.887 & 2                  & $+2$ &  8.511e-16  & '1.922e-01' &  8.115e-16  &  2.390e-16  &  4.318e-16  &  5.357e-16  &  1.021e-15  &  5.484e-16  \\
      14 & 1.887 & 2                  & $-2$ &  5.592e-16  &  4.027e-16  &  5.619e-16  &  1.017e-15  & '1.922e-01' &  8.719e-16  &  2.616e-16  &  8.621e-16  \\
      15 & 1.527 & $2{+}\frac{2}{15}$ & $+2$ &  3.370e-16  &  3.141e-02  &  7.219e-16  &  1.422e-16  &  6.289e-16  &  7.121e-16  &  4.933e-16  &  4.516e-16  \\
      16 & 1.348 & $2{+}\frac{2}{15}$ & $-2$ &  6.436e-16  &  4.340e-17  &  8.115e-16  &  4.521e-16  &  7.064e-03  &  1.821e-16  &  5.489e-16  &  1.304e-16  \\
      17 & 1.348 & $2{+}\frac{2}{15}$ & $+2$ &  1.885e-16  &  7.064e-03  &  1.416e-16  &  5.210e-16  &  4.910e-17  &  6.729e-16  &  4.473e-16  &  7.427e-16  \\
      18 & 1.527 & $2{+}\frac{2}{15}$ & $-2$ &  7.152e-16  &  6.147e-16  &  4.650e-16  &  4.886e-16  &  3.141e-02  &  3.314e-16  &  1.340e-16  &  7.226e-16  \\
    \bottomrule
  \end{tabular}
  \caption{\M{\phi_\beta} and \MB{\phi_\beta}}
  \label{tab:L-p2}
\end{table}

\endgroup

\end{document}